# Thickness-Dependent Differential Reflectance Spectra of Monolayer and Few-Layer MoS$_2$, MoSe$_2$, WS$_2$ and WSe$_2$

Yue Niu [1,2,3], Sergio Gonzalez-Abad [2], Riccardo Frisenda [4], Philipp Marauhn [5], Matthias Drüppel [5], Patricia Gant [4], Robert Schmidt [6], Najme S. Taghavi [4,7], David Barcons [2], Aday J. Molina-Mendoza [8], Steffen Michaelis de Vasconcellos [6], Rudolf Bratschitsch [6], David Perez De Lara [2], Michael Rohlfing [5] and Andres Castellanos-Gomez [4,*]

[1] National Center for International Research on Green Optoelectronics & Guangdong Provincial Key Laboratory of Optical Information Materials and Technology, Institute of Electronic Paper Displays, South China Academy of Advanced Optoelectronics, South China Normal University, Guangzhou 510006, P. R. China.;
[2] Instituto Madrileño de Estudios Avanzados en Nanociencia (IMDEA Nanociencia), Campus de Cantoblanco, E-28049 Madrid, Spain;
[3] National Key Laboratory of Science and Technology on Advanced Composites in Special Environments, Harbin Institute of Technology, Harbin 150001, China
[4] Materials Science Factory, Instituto de Ciencia de Materiales de Madrid (ICMM), Consejo Superior de Investigaciones Científicas (CSIC), Sor Juana Inés de la Cruz 3, 28049 Madrid, Spain;
[5] Institute of Solid-state Theory, University of Münster, 48149 Münster, Germany;
[6] Institute of Physics and Center for Nanotechnology, University of Münster, 48149 Münster, Germany;
[7] Faculty of Physics, Khaje Nasir Toosi University of Technology (KNTU), Tehrān 19697 64499, Iran
[8] Institute of Photonics, Vienna University of Technology, Gusshausstrasse 27–29, 1040 Vienna, Austria;

\* Correspondence: andres.castellanos@csic.es

**Abstract:** The research field of two dimensional (2D) materials strongly relies on optical microscopy characterization tools to identify atomically thin materials and to determine their number of layers. Moreover, optical microscopy-based techniques opened the door to study the optical properties of these nanomaterials. We presented a comprehensive study of the differential reflectance spectra of 2D semiconducting transition metal dichalcogenides (TMDCs), MoS$_2$, MoSe$_2$, WS$_2$, and WSe$_2$, with thickness ranging from one layer up to six layers. We analyzed the thickness-dependent energy of the different excitonic features, indicating the change in the band structure of the different TMDC materials with the number of layers. Our work provided a route to employ differential reflectance spectroscopy for determining the number of layers of MoS$_2$, MoSe$_2$, WS$_2$, and WSe$_2$.

**1. Introduction**

The isolation of atomically thin semiconducting transition metal dichalcogenides (TMDCs) by mechanical exfoliation of bulk layered crystals, aroused the interest of the nanoscience and nanotechnology community on these 2D semiconductors [1–7]. These materials have a band gap within the visible part of the spectrum, bridging the gap between graphene (zero-gap semiconductor) and hexagonal boron nitride (wide-gap semiconductor). Recently, the band gap of semiconductor TMDCs has been exploited to fabricate optoelectronic devices, such as photodetectors [8–14] and solar cells [15–19]. Photoluminescence studies also demonstrated that a reduction in thickness has a strong effect on the band structure of $MoS_2$ and other semiconductor TMDCs, including a change in the band gap and a thickness mediated direct-to-indirect band gap crossover [20–24]. The thickness dependent band gap can have a strong influence on other electrical [25] and optical properties, such as the absorption [25–27], and it has been also exploited to fabricate photodetectors where their spectral bandwidth is determined by the number of layers of the semiconductor channel [9,11]. However, the determination of the intrinsic quantum efficiency and the photoresponse of photodetectors based on semiconducting TMDCs, requires a comprehensive study of their reflectance and/or transmittance with different numbers of layers in a wide spectral range, which is still lacking.

We systematically study the differential reflectance of single- and few-layer $MoS_2$, $MoSe_2$, $WS_2$, and $WSe_2$, from the near-infrared (1.4 eV) to the near-ultraviolet (3.0 eV). The differential reflectance spectra show prominent features due to excitons, and the thickness dependence of these excitonic features is analyzed.

**2. Materials and Method**

We prepared $MoS_2$, $MoSe_2$, $WS_2$, and $WSe_2$ nanosheets by mechanical exfoliation with blue Nitto tape (Nitto Denko Co., Tokyo, Japan, SPV 224P), on commercially available polydimethylsiloxane substrates (Gel-Film from Gelpak®, Hayward, CA, USA). $MoSe_2$, $WS_2$, and $WSe_2$ bulk crystals were synthetic (grown by the vapor transport method), with approximate dimensions of $10 \times 10 \times 0.2$ mm$^3$, whilst the $MoS_2$ material employed in this work was a naturally occurring molybdenite crystal (Moly Hill mine, Quebec, Canada), with approximate dimensions of $50 \times 50 \times 3$ mm$^3$. All the materials studied in the main text were 2H polytype. In the supporting information, we also include results obtained for mechanically exfoliated natural 3R-$MoS_2$ (Mont St. Hilaire, Quebec, Canada), which occurs in the form of micro-crystals, with an area of $0.5 \times 0.5$ mm$^2$ on the surface of a quartz mineral. Few-layer flakes were identified under an optical microscope (Nikon Eclipse CI, Tokyo, Japan), and the number of layers was determined by their opacity in transmission mode. The optical properties of the nanosheets were studied with a home-built micro-reflectance/transmittance setup, described in detail in Reference [28].

The calculations of the absorption spectra were conducted using the *GW*-BSE method, within the LDA + *GdW* approximation [29]. The dielectric screening was implemented by an atom-resolved model function, based on the random phase approximation. For the structural parameters, we used experimental values, as reported in Reference [30] (with $a = 3.160$ Å for $MoS_2$ and $a = 3.299$ Å for $MoSe_2$) and Reference [31] (with $a = 3.155$ Å for $WS_2$ and $a = 3.286$ Å for $WSe_2$). We started with a density functional theory (DFT) calculation within the local density approximation (LDA), using a basis set of localized Gaussian orbitals and norm-conserving pseudopotentials, which also included spin-orbit interaction. The resulting wave functions and energies were used for a subsequent *GdW* calculation, fully considering spin-orbit interaction. For the BSE calculations, we used a 24 × 24 × 1 *k*-point grid for the mono- and bi-layers, and an 18 × 18 × 3 *k*-point grid for the bulk crystals. Notably, we used identical meshes for both the quasiparticle corrections and the electron-hole interaction, so no interpolation scheme was needed. The number of valence and conduction bands in the BSE Hamiltonian were doubled, when going from the monolayer (four/six) to the bilayer and bulk crystals (eight/twelve). A detailed analysis of the convergence of the presented calculation is found in the Supporting Information. For all absorption spectra, an artificial broadening of 35 meV was applied.

## 3. Results and Discussion

Single- and few-layer $MoS_2$, $MoSe_2$, $WS_2$, and $WSe_2$ samples were fabricated by mechanical exfoliation of bulk layered crystals onto a polydimethylsiloxane (PDMS) substrate (Gelfilm from Gelpak®, Hayward, CA, USA, as described in Reference [32]; see the Supporting Information for results obtained with other substrates). We refer the reader to the Materials and Methods section, for more details about the sample fabrication. Moreover, all results shown in the main text were obtained for the 2H polytype, which is the most common in this family of materials. We refer the reader to the Supporting Information, for a comparison between the 2H- and 3R-$MoS_2$ polytypes.

Figure 1a shows a transmission mode optical microscopy image of an exfoliated $MoS_2$ flake, displaying regions with different numbers of layers, as determined from the position of the $E_{2g}$ and $A_{1g}$ lines in their Raman spectra (Figure 1b) [33,34]. Similar optical microscopy images of $MoSe_2$, $WS_2$, and $WSe_2$ samples can be found in the Supporting Information. The quantitative analysis of the red, green, and blue channels of the transmission mode optical images, has recently been proven to be an effective alternative way to determine the number of layers of TMDCs (see description in Reference [35] and Supporting Information in Reference [36]). Figure 1c shows the transmittance $T/T_0$ ($T$: intensity of the light transmitted through the flake, $T_0$: intensity of the light transmitted through the substrate), extracted from the different regions of the transmission mode optical image shown in Figure 1a. The blue channel showed the largest thickness dependence. Thus, it can be very useful to determine the number of layers. The blue channel transmittance dropped monotonically by ~9% per $MoS_2$ layer, in good agreement with the results reported in Reference [35]. We statistically analyzed the blue channel transmittance of 200 $MoS_2$ flakes to get insight of the flake-to-flake variation, finding a typical value of 2–5%, allowing for the accurate determination of the flake thickness, despite the uncertainty introduced by these flake-to-flake variations. A more comprehensive statistical analysis of the blue channel transmittance in the whole family of TMDCs will be published elsewhere, as it lies out of the scope of the current manuscript.

We also found that the blue channel transmittance showed a strong thickness dependence for other TMDCs studied: $WS_2$, $WSe_2$, or $MoSe_2$ (see the Supporting Information for an analogue of Figure 1c, for these materials). This strong thickness dependence of the blue channel transmittance, might be especially relevant to determine the number of layers of relatively thick $MoS_2$ multilayers, as Raman spectroscopy is only accurate in determining layers thinner than 4 layers. The Raman shift difference between the $E_{2g}$ and $A_{1g}$, quickly saturates for flakes thicker than 4 layers, see Reference [33]. Furthermore, for $WS_2$, $WSe_2$ or $MoSe_2$, it is not trivial to determine the number of layers with Raman spectroscopy, as one might need a high resolution Raman system or a system capable to resolve shear Raman modes occurring at low Raman shifts [37–42] (see the Supporting Information to see the thickness dependent Raman spectra of $WS_2$, $WSe_2$ or $MoSe_2$).

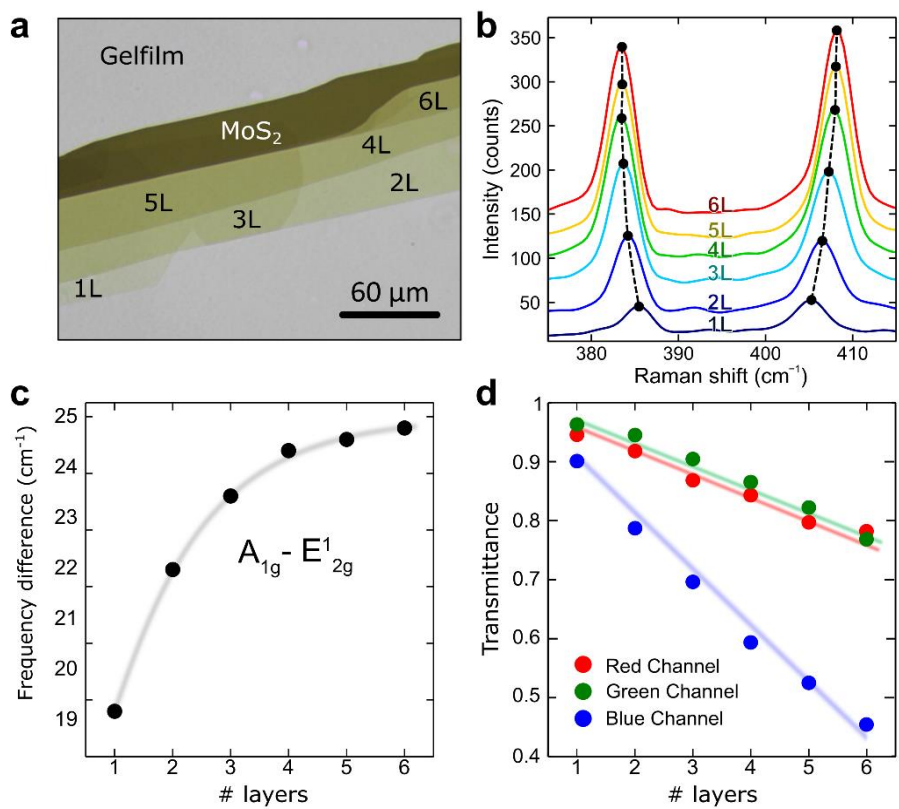

**Figure 1.** (**a**) Transmission mode optical image of a mechanically exfoliated $MoS_2$ flake on polydimethylsiloxane (PDMS) substrate; (**b**) Raman spectra measured on the different regions of the flakes; The thickness of the flake can be determined from the Raman shift difference between the $A_{1g}$ and $E_{2g}$ lines, shown in panel (**c**). Note that a flake-to-flake variation of up to ~0.5 cm$^{-1}$ can be found in the exfoliated flakes, and it would be the main cause of uncertainty in thickness determination through Raman spectroscopy; (**d**) Transmittance of the $MoS_2$ flake (extracted from the red, green, and blue channels of the transmission mode optical images), as a function of the number of layers.

The optical spectra of the fabricated flakes were characterized using a homebuilt micro-reflectance and transmittance setup. We refer the reader to References [28,43], for details on this experimental setup. Briefly, the setup consisted of a Motic BA310 metallurgical microscope equipped with a 50× objective (0.55 numerical aperture and 8.2 mm of working distance), supplemented with a modified trinocular port, which sends part of the reflected light to a multimode fiber-coupled charge coupled device (CCD) spectrometer (Thorlabs CCS200/M, Newton, NJ, USA) to be analyzed. The system can be used to measure differential reflectance and transmittance, with a lateral resolution of ~1 µm. In the Supplementary Information, the reader will find a schematic diagram of the experimental setup configurations used for differential reflectance and transmittance experiments. In the main text, we showed the results of differential reflectance measurements, and we refer the reader to the Supporting Information, for a comparison between differential reflectance and transmittance measurements acquired on the same sample.

The differential reflectance spectrum was calculated as $(R - R_0)/R$, and it is related to the absorption coefficient of the material $\alpha(\lambda)$ as shown in References [26,44]

$$\frac{R - R_0}{R} = \frac{4n}{n_0{}^2 - 1} \alpha(\lambda) \tag{1}$$

where $R$ is the intensity reflected by the flake, $R_0$ the intensity reflected by the substrate, $n$ is the refractive index of the flake under study, and $n_0$ is the refractive index of the substrate. Figure 2 shows the differential reflectance spectra measured on the single- and few-layer regions, for the different semiconductor TMDCs studied. The spectra acquired for this family of 2D materials showed overall similar features: Pronounced peaks corresponding to the generation of excitons. The exact energy at which these peaks appeared is material-to-material dependent, because those features were determined by the band structures of these different compounds. The exciton peaks in Figure 2 are labelled A, B, and C (and D for $WSe_2$), following the nomenclature employed in the literature to name the different excitons in semiconducting TMDCs [20–22]. The A exciton, occurring near the absorption band edge, corresponded to direct band gap transitions at the *K* point in the Brillouin zone [20–22]. This feature is the most studied one, as it is also the dominant one in photoluminescence spectra. Close to the A exciton peak, at slightly higher energy, the transition metal dichalcogenides showed another prominent peak in their differential reflectance spectra, corresponding to another direct band gap transition at the *K* point, but at higher energy, which yielded the creation of the so-called B excitons. For monolayer TMDCs, the origin of this higher energy transition at the *K* point was related to the splitting of the valence band, due to the spin-orbit interaction. For multilayer systems, the splitting of the valence band was driven by a combination of spin-orbit and interlayer interaction.

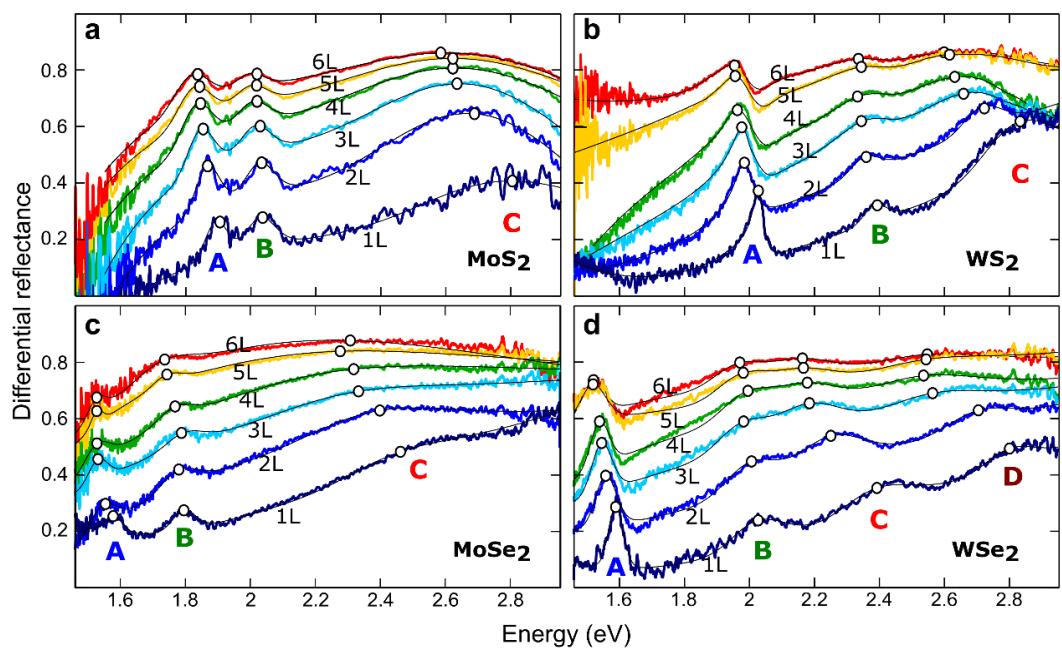

**Figure 2.** Differential reflectance spectra, measured as a function of the number of layers for (**a**) $MoS_2$, (**b**) $WS_2$, (**c**) $MoSe_2$, and (**d**) $WSe_2$. The spectra have been fitted to a sum of Lorentzian/Gaussian peaks (solid thin black lines), to determine the position of the different excitonic features (highlighted by white circles).

Apart from the narrow A and B exciton peaks, the differential reflectance spectra of $MoS_2$, $MoSe_2$, and $WS_2$, also showed other broader spectroscopic features in an energy range from 2.5 eV to 2.9 eV (referred to as the C exciton peak). This was due to singularities in the joint density of states between the first valence and conduction bands, in a circle around the $\Gamma$ point (into the local minimum of the lowest conduction band between $\Gamma$ and $K$), which led to multiple optical transitions to nearly degenerate in energy [26,27,45–49]. Regarding WSe$_2$, we found that instead of just one broad C exciton feature, two features labeled C and D, were reported in recent absorption measurements [49]. Whilst the C exciton of WSe$_2$ consisted of several transitions along the $\Gamma$-$K$ direction, between the highest valence and lowest conduction bands, the D exciton had the largest contributions from the spin-split lower valence band, into the lowest conduction band [49].

The differential reflectance spectra were fit to a sum of Gaussian/Lorentzian peaks, with a broad background to determine the peak position, width, and magnitude of the excitonic features, as a function of the number of layers for the different 2D semiconductor materials. The thin black lines in Figure 2 correspond to the resulting fits for the different measured spectra, and the empty circles highlight the energy values determined for the different excitons from the fits. We found that the flake-to-flake variation in the peak position was the most relevant source for uncertainty, in the analysis of the thickness dependent spectra. In the Supplementary Information, we compare the spectra acquired on 6 single-layers and 4 bilayers of MoS$_2$, to illustrate the typical flake-to-flake variation in the spectra. The exciton positions can shift up to 10–15 meV, and the flake-to-flake variations in intensity can reach 5–10%.

To visualize the thickness dependence of the exciton energies, Figure 3 summarizes the determined exciton energies, as a function of the number of layers for MoS$_2$, MoSe$_2$, WS$_2$, and WSe$_2$. The A exciton peak redshifts as the thickness increases for all the studied materials, in agreement with previous photoluminescence results. The B exciton displayed a non-monotonic thickness dependence, which might arise from the moderate thickness dependence and the 10–15 meV flake-to-flake variation of the peak position. Overall, the B exciton feature also shifted to lower energy with the increasing number of layers. As discussed above, for single layers, the separation between the A and B exciton peaks is due to the spin-orbit splitting of the valence band. Therefore, the larger spin-orbit splitting induced by the heavier W atoms with respect to Mo atoms, is translated to a larger separation of the A and B features in the differential reflectance spectra of W-based TMDCs. Moreover, Se- based dichalcogenides exhibited a larger splitting between the A and B exciton peaks, than that of S- based dichalcogenides. Table 1 summarizes the values of the splitting between the A and B excitons for the single-layer TMDCs studied in this work and compares these values with theoretical values obtained through *ab initio* calculations (see the Supporting Information for more details about the calculations). The good agreement between predicted A-B splitting and the experimental values, confirms our theoretical calculations captured the essential features to describe the excitonic properties of TMDCs. Note that experimental values for the A-B splitting reported in the literature show variability [49–51], e.g., due to different experimental techniques and flake-to-flake variations due to small differences in strain, doping, and/or presence of defects. This explains the variation in discrepancies between our predicted and experimental AB-splittings.

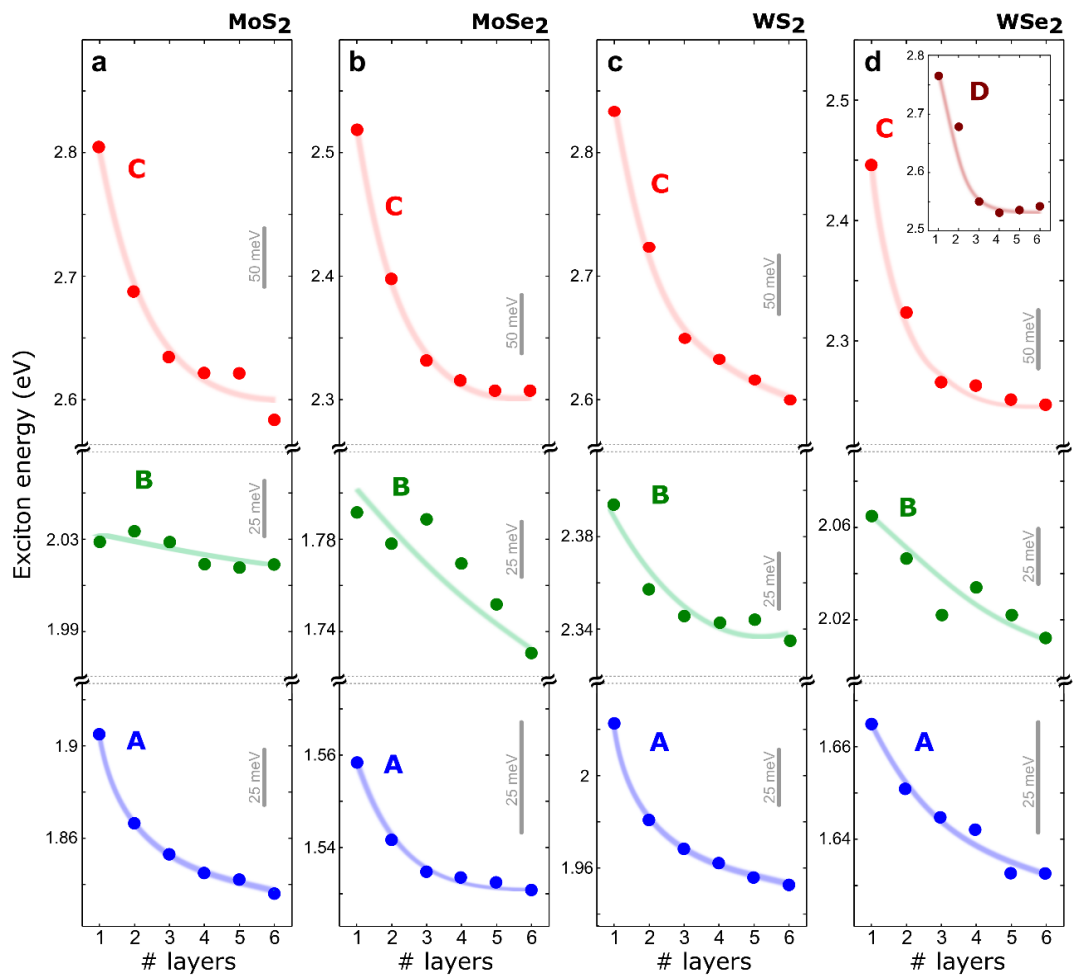

**Figure 3.** Thickness dependence of the exciton energies, extracted from the measured differential reflectance spectra of (**a**) $MoS_2$, (**b**) $WS_2$, (**c**) $MoSe_2$, and (**d**) $WSe_2$. The solid lines are guides to the eye.

**Table 1.** Comparison of the spin-orbit splitting extracted from the differential reflectance spectra, and those obtained from ab initio calculations, including spin-orbit interaction.

| Material | Experimental A-B splitting (meV) | Theoretical A-B splitting (meV) |
|---|---|---|
| 1L – $MoS_2$ | 124 ± 5 | 152 |
| 1L – $MoSe_2$ | 219 ± 10 | 218 |
| 1L – $WS_2$ | 371 ± 5 | 420 |
| 1L – $WSe_2$ | 398 ± 10 | 464 |

Interestingly, we also found that the C exciton showed a prominent shift with the thickness, even more pronounced than that of the A exciton. Note that the number of works studying the C excitonic feature are still very scarce, as most experiments employ photoluminescence with green laser excitation ($E$ ~2 eV–2.3 eV) to observe the generated excitons.

## 4. Conclusions

In summary, we presented a systematic study of the differential reflectance spectra $MoS_2$, $MoSe_2$, $WS_2$, and $WSe_2$, from the near-infrared (1.4 eV) to the near-ultraviolet (3.0 eV). The differential reflectance spectra showed prominent features due to the generation of excitons, and the energy at which these features appear depends on the thickness of the flakes, because of quantum confinement effects. We proposed employing a combination of a quantitative analysis of transmission mode optical images and differential reflectance measurements, as an alternative method to determine the number of layers.

**Supplementary Materials:** The following are available online at www.mdpi.com/xxx/s1, Figure S1: Transmission mode optical images of mechanically exfoliated transition metal dichalcogenides (TMDCs) onto polydimethylsiloxane (PDMS) substrates; Figure S2: Transmittance (extracted from the red, green, and blue channels of the transmission mode optical images), as a function of the number of layers; Figure S3: Blue channel transmittance measured on more than 200 $MoS_2$ flakes with different numbers of layers (ranging from 1 layer to 4 layers); Figure S4: Comparison between the differential reflectance spectra measured for single-layer $MoS_2$ on different substrates: glass, polycarbonate (PC), polypropylene (PP), and polydimethylsiloxane (PDMS); Figure S5: Reflection mode optical images of $MoS_2$, $WS_2$, $MoSe_2$, and $WSe_2$ after transfer onto $SiO_2$/Si substrates (285 nm thick $SiO_2$); Figure S6: Thickness dependence of the optical contrast measured for $MoS_2$, $WS_2$, $MoSe_2$, and $WSe_2$ deposited onto $SiO_2$/Si substrates (285 nm thick $SiO_2$); Figure S7: Raman spectra measured on $MoS_2$, $WS_2$ , $MoSe_2$, and $WSe_2$ deposited onto PDMS substrates; Figure S8: Quantitative analysis of the Raman spectra measured on $MoS_2$ , $WS_2$ , $MoSe_2$, and $WSe_2$ deposited onto PDMS substrates; Figure S9: Comparison between differential reflectance and transmittance measurements, carried out on the same $MoS_2$ flakes on PDMS; Figure S10: Differential reflectance intensity, measured from the differential reflectance spectra shown in Figure S6, at energies outside the excitonic resonance windows; Figure S11: Comparison between different methods to measure the optical properties of 2D materials (using 1L, 2L, and 3L $MoS_2$ as testbeds); Figure S12: Comparison between the crystal structure of the 2H- and 3R- polytypes; Figure S13: Transmission mode optical images of 2H-$MoS_2$ and 3R-$MoS_2$ on PDMS; Figure S14: Transmittance (extracted from the red, green, and blue channels of the transmission mode optical images), as a function of the number of layers for 2H-$MoS_2$ and 3R-$MoS_2$; Figure S15: Differential reflectance spectra, measured as a function of the number of layers for 2H- and 3R-$MoS_2$; Figure S16: Thickness dependence of the exciton energies, extracted from the differential reflectance spectra of 2H-$MoS_2$ and 3R-$MoS_2$; Figure S17: Convergence of the quasiparticle gap at the *K* point, with respect to the energy cutoff used in the LDA + *GdW* approach, for the representation of ε and *W*; Figure S18: Convergence of the A and B exciton for all four TMDCs, with respect to the *k*-point grid used in the BSE; Figure S19: Convergence of the A and B exciton for the bulk crystal of $MoS_2$, with respect to the *k*-point grid applied in the BSE; Figure S20: Absorption spectra of $MoS_2$, $WS_2$, $MoSe_2$, and $WSe_2$, for all studied number of layers; Figure S21: Direct comparison between the exciton energies obtained experimentally, and those calculated for $MoS_2$, $WS_2$, $MoSe_2$, and $WSe_2$; Figure S22: Comparison between differential reflectance spectra acquired on different single-layer and bilayer $MoS_2$ flakes, which illustrates how the flake-to-flake variation is the main source of uncertainty in the analysis of the thickness dependent spectra of TMDCs; Figure S23: Schematic of the experimental setup employed for the micro-reflectance and transmittance measurements on TMDCs.

**Author Contributions:** Y.N., S.G.-A., P.G., A.J.M.-M., R.F., N.S.T., D.B., D.P.D.L. and A.C.-G. fabricated the samples, performed the optical spectroscopy measurements and analyzed the experimental results. R.S., S.M.d.V. and R.B. provided the bulk $MoSe_2$, $WS_2$ and $WSe_2$ crystals. P.M., M.D. and M.R. performed the ab-initio calculations. All the authors discussed the results and contributed in the elaboration of the manuscript.

**Funding:** This project received funding from the European Research Council (ERC) under the European Union's Horizon 2020 research, innovation programme (grant agreement n° 755655, ERC-StG 2017 project 2D-TOPSENSE), the EU Graphene Flagship funding (Grant Graphene Core 2, 785219), the Netherlands Organisation for Scientific Research (NWO) through the research program Rubicon with project number 680-50-1515, the MINECO (program FIS2015-67367-C2-1-P), and the China Scholarship Council (File NO. 201506120102).

**Acknowledgment:** We thank Professor Emilio M. Pérez for his support with the Raman spectroscopy measurements. Andres Castellanos-Gomez and Patricia Gant acknowledge funding from the EU Graphene Flagship funding (Grant Graphene Core 2, 785219). Riccardo Frisenda acknowledges support from the Netherlands Organisation for Scientific Research (NWO), through the research program Rubicon with project number 680-50-1515. David Perez De Lara acknowledges support from the MINECO (program FIS2015-67367-C2-1-P). Yue Niu acknowledges the grant from the China Scholarship Council (File NO. 201506120102). Philipp Marauhn, Matthias Drüppel, and Michael Rohlfing gratefully acknowledge the computing time granted by the John von Neumann Institute for Computing (NIC), and provided on the super-computer JURECA at Jülich Supercomputing Centre (JSC)

**References**

1. Novoselov, K.S.; Jiang, D.; Schedin, F.; Booth, T.J.; Khotkevich, V.V.; Morozov, S.V.; Geim, A.K. Two-dimensional atomic crystals. *Proc. Natl. Acad. Sci. USA* **2005**, *102*, 10451–10453, doi:10.1073/pnas.0502848102.
2. Xu, M.; Liang, T.; Shi, M.; Chen, H. Graphene-like two-dimensional materials. *Chem. Rev.* **2013**, *113*, 3766–3798, doi:10.1021/cr300263a.
3. Xia, F.; Wang, H.; Xiao, D.; Dubey, M.; Ramasubramaniam, A. Two-dimensional material nanophotonics. *Nat. Photonics* **2014**, *8*, 899–907, doi:10.1038/nphoton.2014.271.
4. Lv, R.; Robinson, J.A.; Schaak, R.E.; Sun, D.; Sun, Y.; Mallouk, T.E.; Terrones, M. Transition metal dichalcogenides and beyond: synthesis, properties, and applications of single- and few-layer nanosheets. *Acc. Chem. Res.* **2015**, *48*, 56–64, doi:10.1021/ar5002846.
5. Wang, Q.H.; Kalantar-Zadeh, K.; Kis, A.; Coleman, J.N.; Strano, M.S. Electronics and optoelectronics of two-dimensional transition metal dichalcogenides. *Nat. Nanotechnol.* **2012**, *7*, 699–712, doi:10.1038/nnano.2012.193.
6. Castellanos-Gomez, A. Why all the fuss about 2D semiconductors? *Nat. Photonics* **2016**, *10*, 202–204, doi:10.1038/nphoton.2016.53.
7. Avouris, P.; Xia, F. Graphene applications in electronics and photonics. *MRS Bull.* **2012**, *37*, 1225–1234, doi:10.1557/mrs.2012.206.
8. Yin, Z.; Li, H.; Li, H.; Jiang, L.; Shi, Y.; Sun, Y.; Lu, G.; Zhang, Q.; Chen, X.; Zhang, H. Single-layer $MoS_2$ phototransistors. *ACS Nano* **2012**, *6*, 74–80, doi:10.1021/nn2024557.
9. Lee, H.S.; Min, S.-W.; Chang, Y.-G.; Park, M.K.; Nam, T.; Kim, H.; Kim, J.H.; Ryu, S.; Im, S. $MoS_2$ nanosheet phototransistors with thickness-modulated optical energy gap. *Nano Lett.* **2012**, *12*, 3695–3700, doi:10.1021/nl301485q.
10. Zhang, W.; Chiu, M.-H.; Chen, C.-H.; Chen, W.; Li, L.-J.; Wee, A.T.S. Role of metal contacts in high-performance phototransistors based on $WSe_2$ monolayers. *ACS Nano* **2014**, *8*, 8653–8661, doi:10.1021/nn503521c.
11. Choi, W.; Cho, M.Y.; Konar, A.; Lee, J.H.; Cha, G.-B.; Hong, S.C.; Kim, S.; Kim, J.; Jena, D.; Joo, J.; et al. High-detectivity multilayer $MoS_2$ phototransistors with spectral response from ultraviolet to infrared. *Adv. Mater.* **2012**, *24*, 5832–5836, doi:10.1002/adma.201201909.
12. Zhang, W.; Huang, J.-K.; Chen, C.-H.; Chang, Y.-H.; Cheng, Y.-J.; Li, L.-J. High-gain phototransistors based on a CVD $MoS_2$ monolayer. *Adv. Mater.* **2013**, *25*, 3456–3461, doi:10.1002/adma.201301244.
13. Abderrahmane, A.; Ko, P.J.; Thu, T.V.; Ishizawa, S.; Takamura, T.; Sandhu, A. High photosensitivity few-layered $MoSe_2$ back-gated field-effect phototransistors. *Nanotechnology* **2014**, *25*, 365202, doi:10.1088/0957-4484/25/36/365202.
14. Lopez-Sanchez, O.; Lembke, D.; Kayci, M.; Radenovic, A.; Kis, A. Ultrasensitive photodetectors based on monolayer $MoS_2$. *Nat. Nanotechnol.* **2013**, *8*, 497–501, doi:10.1038/nnano.2013.100.
15. Ross, J.S.; Klement, P.; Jones, A.M.; Ghimire, N.J.; Yan, J.; Mandrus, D.G.; Taniguchi, T.; Watanabe, K.; Kitamura, K.; Yao, W.; et al. Electrically tunable excitonic light-emitting diodes based on monolayer $WSe_2$ p-n junctions. *Nat. Nanotechnol.* **2014**, *9*, 268–272, doi:10.1038/nnano.2014.26.
16. Baugher, B.W.H.; Churchill, H.O.H.; Yang, Y.; Jarillo-Herrero, P. Optoelectronic devices based on electrically tunable p-n diodes in a monolayer dichalcogenide. *Nat. Nanotechnol.* **2014**, *9*, 262–267, doi:10.1038/nnano.2014.25.
17. Pospischil, A.; Furchi, M.M.; Mueller, T. Solar-energy conversion and light emission in an atomic monolayer p-n diode. *Nat. Nanotechnol.* **2014**, *9*, 257–261, doi:10.1038/nnano.2014.14.
18. Lee, C.-H.; Lee, G.-H.; van der Zande, A.M.; Chen, W.; Li, Y.; Han, M.; Cui, X.; Arefe, G.; Nuckolls, C.; Heinz, T.F.; et al. Atomically thin p–n junctions with van der Waals heterointerfaces. *Nat. Nanotechnol.* **2014**, *9*, 676–681, doi:10.1038/nnano.2014.150.
19. Groenendijk, D.J.; Buscema, M.; Steele, G.A.; Michaelis de Vasconcellos, S.; Bratschitsch, R.; van der Zant, H.S.J.; Castellanos-Gomez, A. Photovoltaic and photothermoelectric effect in a double-gated $WSe_2$ device. *Nano Lett.* **2014**, *14*, 5846–5852, doi:10.1021/nl502741k.
20. Splendiani, A.; Sun, L.; Zhang, Y.; Li, T.; Kim, J.; Chim, C.-Y.; Galli, G.; Wang, F. Emerging photoluminescence in monolayer $MoS_2$. *Nano Lett.* **2010**, *10*, 1271–1275, doi:10.1021/nl903868w.

21. Mak, K.F.; Lee, C.; Hone, J.; Shan, J.; Heinz, T.F. Atomically Thin MoS$_2$: A new direct-gap semiconductor. *Phys. Rev. Lett.* **2010**, *105*, 136805, doi:10.1103/PhysRevLett.105.136805.
22. Zhao, W.; Ghorannevis, Z.; Chu, L.; Toh, M.; Kloc, C.; Tan, P.-H.; Eda, G. Evolution of electronic structure in atomically thin sheets of WS$_2$ and WSe$_2$. *ACS Nano* **2013**, *7*, 791–797, doi:10.1021/nn305275h.
23. Li, Y.; Li, X.; Yu, T.; Yang, G.; Chen, H.; Zhang, C.; Feng, Q.; Ma, J.; Liu, W.; Xu, H. Accurate identification of layer number for few-layer WS$_2$ and WSe$_2$ via spectroscopic study. *Nanotechnology* **2018**, *29*, 124001.
24. Ottaviano, L.; Palleschi, S.; Perrozzi, F.; D'Olimpio, G.; Priante, F.; Donarelli, M.; Benassi, P.; Nardone, M.; Gonchigsuren, M.; Gombosuren, M. Mechanical exfoliation and layer number identification of MoS$_2$ revisited. *2D Mater.* **2017**, *4*, 45013.
25. Li, D.; Song, X.; Xu, J.; Wang, Z.; Zhang, R.; Zhou, P.; Zhang, H.; Huang, R.; Wang, S.; Zheng, Y. Optical properties of thickness-controlled MoS$_2$ thin films studied by spectroscopic ellipsometry. *Appl. Surf. Sci.* **2017**, *421*, 884–890.
26. Dhakal, K.P.; Duong, D.L.; Lee, J.; Nam, H.; Kim, M.; Kan, M.; Lee, Y.H.; Kim, J. Confocal absorption spectral imaging of MoS$_2$: Optical transitions depending on the atomic thickness of intrinsic and chemically doped MoS$_2$. *Nanoscale* **2014**, *6*, 13028–13035, doi:10.1039/c4nr03703k.
27. Castellanos-Gomez, A.; Quereda, J.; van der Meulen, H.P.; Agraït, N.; Rubio-Bollinger, G. Spatially resolved optical absorption spectroscopy of single- and few-layer MoS$_2$ by hyperspectral imaging. *Nanotechnology* **2016**, *27*, 115705, doi:10.1088/0957-4484/27/11/115705.
28. Frisenda, R.; Niu, Y.; Gant, P.; Molina-Mendoza, A.J.; Schmidt, R.; Bratschitsch, R.; Liu, J.; Fu, L.; Dumcenco, D.; Kis, A.; et al. Micro-reflectance and transmittance spectroscopy: A versatile and powerful tool to characterize 2D materials. *J. Phys. D Appl. Phys.* **2017**, *50*, 74002.
29. Rohlfing, M. Electronic excitations from a perturbative LDA + GdW approach. *Phys. Rev. B* **2010**, *82*, 205127, doi:10.1103/PhysRevB.82.205127.
30. Böker, T.; Severin, R.; Müller, A.; Janowitz, C.; Manzke, R.; Voß, D.; Krüger, P.; Mazur, A.; Pollmann, J. Band structure of MoS$_2$, MoSe$_2$, and α-MoTe$_2$: Angle-resolved photoelectron spectroscopy and *ab initio* calculations. *Phys. Rev. B* **2001**, *64*, 235305, doi:10.1103/PhysRevB.64.235305.
31. Yun, W.S.; Han, S.W.; Hong, S.C.; Kim, I.G.; Lee, J.D. Thickness and strain effects on electronic structures of transition metal dichalcogenides: 2H-$MX_2$ semiconductors ($M$ = Mo, W; $X$ = S, Se, Te). *Phys. Rev. B* **2012**, *85*, 33305, doi:10.1103/PhysRevB.85.033305.
32. Buscema, M.; Steele, G.A.; van der Zant, H.S.J.; Castellanos-Gomez, A. The effect of the substrate on the Raman and photoluminescence emission of single-layer MoS$_2$. *Nano Res.* **2014**, *7*, 561–571, doi:10.1007/s12274-014-0424-0.
33. Lee, C.; Yan, H.; Brus, L.E.; Heinz, T.F.; Hone, Ḱ.J.; Ryu, S. Anomalous Lattice Vibrations of Single-and Few-Layer MoS$_2$. *ACS Nano* **2010**, *4*, 2695–2700.
34. Placidi, M.; Dimitrievska, M.; Izquierdo-Roca, V.; Fontané, X.; Castellanos-Gomez, A.; Pérez-Tomás, A.; Mestres, N.; Espindola-Rodriguez, M.; López-Marino, S.; Neuschitzer, M. Multiwavelength excitation Raman scattering analysis of bulk and two-dimensional MoS$_2$: Vibrational properties of atomically thin MoS$_2$ layers. *2D Mater.* **2015**, *2*, 35006.
35. Zhang, H.; Ran, F.; Shi, X.; Fang, X.; Wu, S.; Liu, Y.; Zheng, X.; Yang, P.; Liu, Y.; Wang, L.; et al. Optical thickness identification of transition metal dichalcogenide nanosheets on transparent substrates. *Nanotechnology* **2017**, *28*, 164001, doi:10.1088/1361-6528/aa6133.
36. Castellanos-Gomez, A.; Roldán, R.; Cappelluti, E.; Buscema, M.; Guinea, F.; van der Zant, H.S.J.; Steele, G.A. Local strain engineering in atomically thin MoS$_2$. *Nano Lett.* **2013**, *13*, 5361–5366, doi:10.1021/nl402875m.
37. Plechinger, G.; Heydrich, S.; Eroms, J.; Weiss, D.; Schüller, C.; Korn, T. Raman spectroscopy of the interlayer shear mode in few-layer MoS$_2$ flakes. *Appl. Phys. Lett.* **2012**, *101*, 101906.
38. Zhao, W.; Ghorannevis, Z.; Amara, K.K.; Pang, J.R.; Toh, M.; Zhang, X.; Kloc, C.; Tan, P.H.; Eda, G. Lattice dynamics in mono-and few-layer sheets of WS$_2$ and WSe$_2$. *Nanoscale* **2013**, *5*, 9677–9683.
39. Zhang, X.; Han, W.P.; Wu, J.B.; Milana, S.; Lu, Y.; Li, Q.Q.; Ferrari, A.C.; Tan, P.H. Raman spectroscopy of shear and layer breathing modes in multilayer MoS$_2$. *Phys. Rev. B* **2013**, *87*, 115413.
40. Zhao, Y.; Luo, X.; Li, H.; Zhang, J.; Araujo, P.T.; Gan, C.K.; Wu, J.; Zhang, H.; Quek, S.Y.; Dresselhaus, M.S. Interlayer breathing and shear modes in few-trilayer MoS$_2$ and WSe$_2$. *Nano Lett.* **2013**, *13*, 1007–1015.
41. Tonndorf, P.; Schmidt, R.; Böttger, P.; Zhang, X.; Börner, J.; Liebig, A.; Albrecht, M.; Kloc, C.; Gordan, O.; Zahn, D.R.T.; et al. Photoluminescence emission and Raman response of monolayer MoS$_2$, MoSe$_2$, and WSe$_2$. *Opt. Express* **2013**, *21*, 4908, doi:10.1364/OE.21.004908.

42. Puretzky, A.A.; Liang, L.; Li, X.; Xiao, K.; Wang, K.; Mahjouri-Samani, M.; Basile, L.; Idrobo, J.C.; Sumpter, B.G.; Meunier, V. Low-frequency Raman fingerprints of two-dimensional metal dichalcogenide layer stacking configurations. *ACS Nano* **2015**, *9*, 6333–6342.
43. Ghasemi, F.; Frisenda, R.; Dumcenco, D.; Kis, A.; Perez de Lara, D.; Castellanos-Gomez, A. High Throughput Characterization of Epitaxially Grown Single-Layer $MoS_2$. *Electronics* **2017**, *6*, 28.
44. McIntyre, J.D.E.; Aspnes, D.E. Differential reflection spectroscopy of very thin surface films. *Surf. Sci.* **1971**, *24*, 417–434, doi:10.1016/0039-6028(71)90272-X.
45. Qiu, D.Y.; da Jornada, F.H.; Louie, S.G. Optical Spectrum of $MoS_2$: Many-body effects and diversity of exciton states. *Phys. Rev. Lett.* **2013**, *111*, 216805, doi:10.1103/PhysRevLett.111.216805.
46. Kozawa, D.; Kumar, R.; Carvalho, A.; Kumar Amara, K.; Zhao, W.; Wang, S.; Toh, M.; Ribeiro, R.M.; Castro Neto, A.H.; Matsuda, K.; et al. Photocarrier relaxation pathway in two-dimensional semiconducting transition metal dichalcogenides. *Nat. Commun.* **2014**, *5*, 193–335, doi:10.1038/ncomms5543.
47. Klots, A.R.; Newaz, A.K.M.; Wang, B.; Prasai, D.; Krzyzanowska, H.; Lin, J.; Caudel, D.; Ghimire, N.J.; Yan, J.; Ivanov, B.L.; et al. Probing excitonic states in suspended two-dimensional semiconductors by photocurrent spectroscopy. *Sci. Rep.* **2014**, *4*, 6608, doi:10.1038/srep06608.
48. Gibaja, C.; Rodriguez-San-Miguel, D.; Ares, P.; Gómez-Herrero, J.; Varela, M.; Gillen, R.; Maultzsch, J.; Hauke, F.; Hirsch, A.; Abellán, G.; et al. Few-layer antimonene by liquid-phase exfoliation. *Angew. Chem. Int. Ed.* **2016**, doi:10.1002/anie.201605298.
49. Schmidt, R.; Niehues, I.; Schneider, R.; Drüppel, M.; Deilmann, T.; Rohlfing, M.; de Vasconcellos, S.M.; Castellanos-Gomez, A.; Bratschitsch, R. Reversible uniaxial strain tuning in atomically thin $WSe_2$. *2D Mater.* **2016**, *3*, 21011, doi:10.1088/2053-1583/3/2/021011.
50. Frisenda, R.; Drüppel, M.; Schmidt, R.; Michaelis de Vasconcellos, S.; Perez de Lara, D.; Bratschitsch, R.; Rohlfing, M.; Castellanos-Gomez, A. Biaxial strain tuning of the optical properties of single-layer transition metal dichalcogenides. *NPJ 2D Mater. Appl.* **2017**, *1*, 10, doi:10.1038/s41699-017-0013-7.
51. Arora, A.; Koperski, M.; Nogajewski, K.; Marcus, J.; Faugeras, C.; Potemski, M. Excitonic resonances in thin films of $WSe_2$: From monolayer to bulk material. *Nanoscale* **2015**, *7*, 10421–10429.

**Supporting information:**

# Thickness-Dependent Differential Reflectance Spectra of Monolayer and Few-Layer MoS$_2$, MoSe$_2$, WS$_2$ and WSe$_2$

**Yue Niu** [1,2,3], **Sergio Gonzalez-Abad** [2], **Riccardo Frisenda** [4], **Philipp Marauhn** [5], **Matthias Drüppel** [5], **Patricia Gant** [4], **Robert Schmidt** [6], **Najme S. Taghavi** [4,7], **David Barcons** [2], **Aday J. Molina-Mendoza** [8], **Steffen Michaelis de Vasconcellos** [6], **Rudolf Bratschitsch** [6], **David Perez De Lara** [2], **Michael Rohlfing** [5] **and Andres Castellanos-Gomez** [4,*]

[1] National Center for International Research on Green Optoelectronics & Guangdong Provincial Key Laboratory of Optical Information Materials and Technology, Institute of Electronic Paper Displays, South China Academy of Advanced Optoelectronics, South China Normal University, Guangzhou 510006, P. R. China.;
[2] Instituto Madrileño de Estudios Avanzados en Nanociencia (IMDEA Nanociencia), Campus de Cantoblanco, E-28049 Madrid, Spain;
[3] National Key Laboratory of Science and Technology on Advanced Composites in Special Environments, Harbin Institute of Technology, Harbin 150001, China
[4] Materials Science Factory, Instituto de Ciencia de Materiales de Madrid (ICMM), Consejo Superior de Investigaciones Científicas (CSIC), Sor Juana Inés de la Cruz 3, 28049 Madrid, Spain;
[5] Institute of Solid-state Theory, University of Münster, 48149 Münster, Germany;
[6] Institute of Physics and Center for Nanotechnology, University of Münster, 48149 Münster, Germany;
[7] Faculty of Physics, Khaje Nasir Toosi University of Technology (KNTU), Tehrān 19697 64499, Iran
[8] Institute of Photonics, Vienna University of Technology, Gusshausstrasse 27–29, 1040 Vienna, Austria;

* Correspondence: andres.castellanos@csic.es

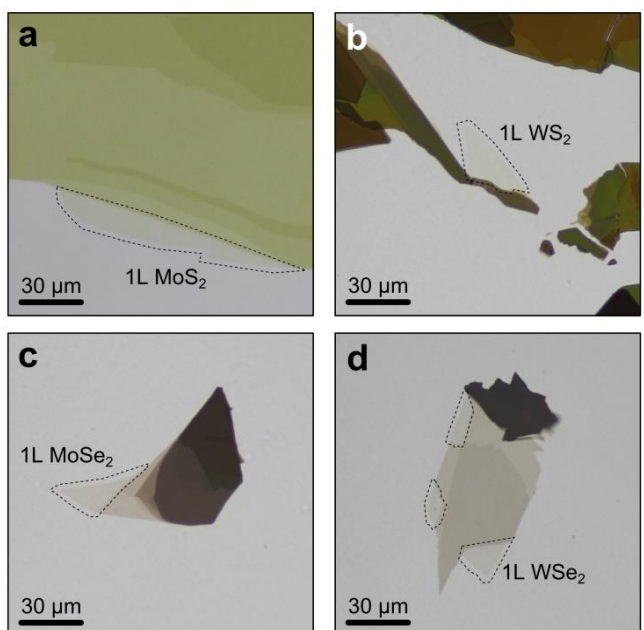

**Figure S1.** Transmission mode optical images of mechanically exfoliated TMDCs onto PDMS substrates. (a) MoS$_2$. (b) WS$_2$. (c) MoSe$_2$. (d) WSe$_2$. Single-layer areas have been highlighted with a dashed black line.

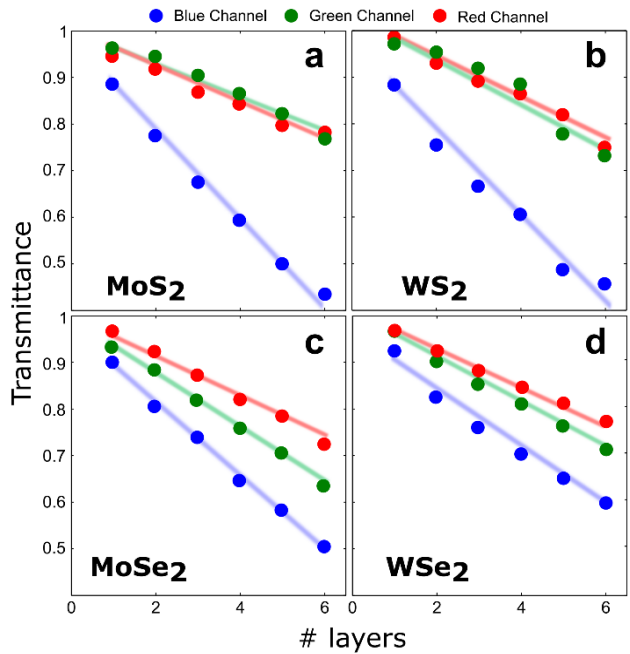

**Figure S2.** Trasmittance (extracted from the red, green and blue channels of the trasnmission mode optical images) as a function of the number of layers. (a) MoS$_2$. (b) WS$_2$. (c) MoSe$_2$. (d) WSe$_2$. The solid lines are guides to the eye.

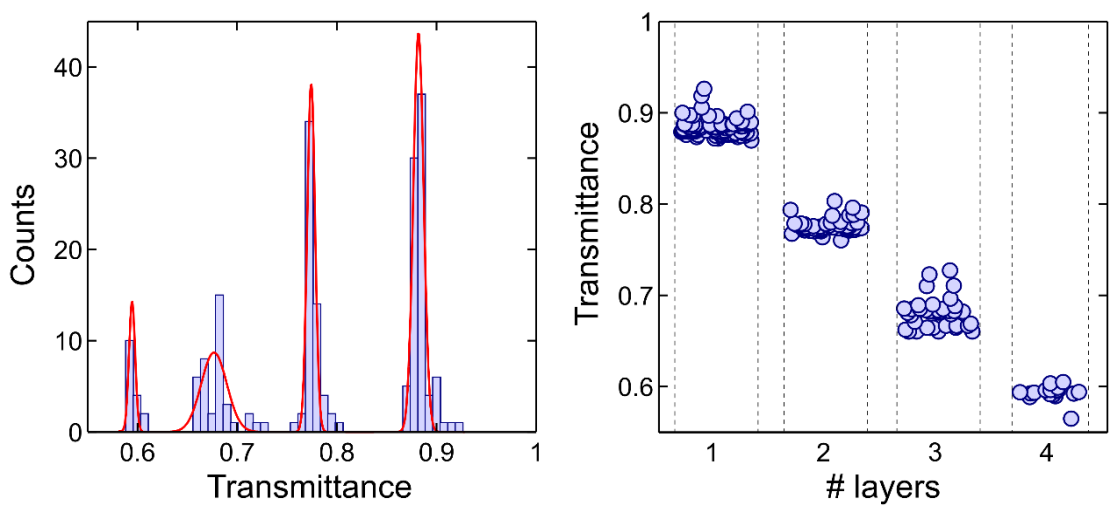

**Figure S3.** (a) Histogram of the blue channel transmittance measured on more than 200 MoS$_2$ flakes with different number of layers (ranging from 1 layer to 4 layers). The histogram has been fitted to a sum of 4 Gaussian curves. (b) Number of layers assigned from the transmittance of the blue channel of the same Mos$_2$ flakes shown in (a).

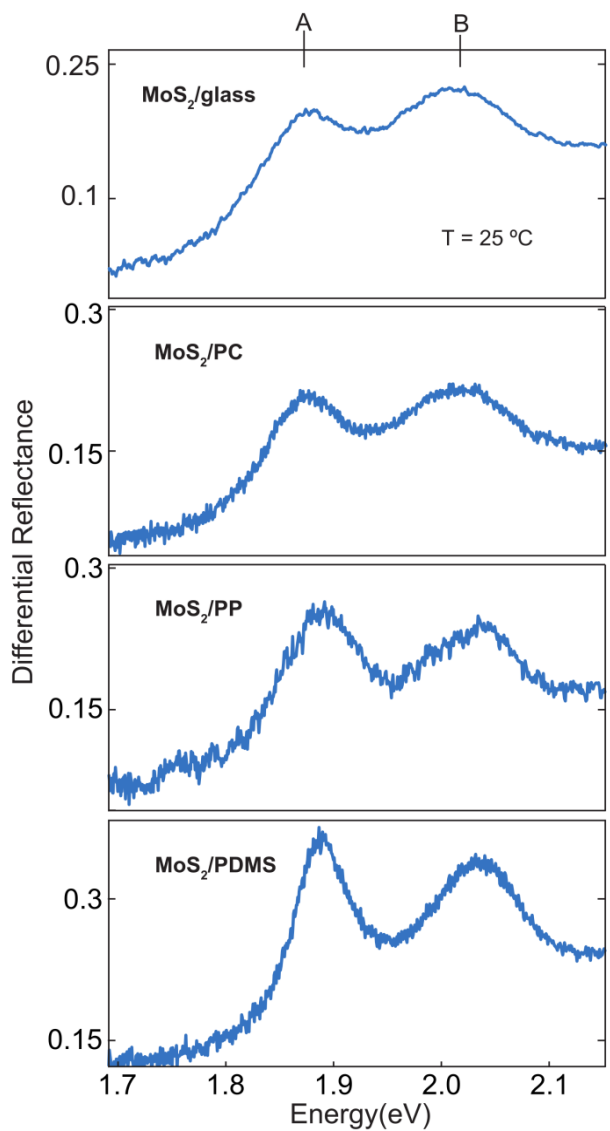

**Figure S4.** Comparison between the differential reflectance spectra measured for single-layer $MoS_2$ on different substrates: glass, polycarbonate (PC), polypropylene (PP) and poly-dimethil siloxane (PDMS).

One advantage of choosing PDMS as substrate for the characterization of TMDCs is that once the flakes are fully characterized they can be easily transferred to another substrate by means of an all-dry transfer method that exploits the viscoelasticity of PDMS to accomplish the transfer of the flake.[45] Figure S5 shows some examples of TMDC flakes that have been transferred from the PDMS substrate to a silicon substrate with a 285 nm $SiO_2$ capping layer, which is one of the standard substrates employed in many laboratories working with graphene and other 2D materials.

For 2D materials supported on $SiO_2$/Si substrates the quantitative analysis of their optical contrast (defined as $C = (I_{flake}-I_{subs})/(I_{flake}+I_{subs})$) is a common method to identify atomically thin flakes and to estimate their number of layers.[46–52] These analyses are typically carried out by acquiring reflection mode optical images while the illumination wavelength is selected by means of narrow bandpass filters[46,49,50], by hyperspectral imaging [39,53], or by using the micro-reflectance setup employed in this work.[32,33] Figure S6 shows a summary of the optical contrast spectra acquired for $MoS_2$, $WS_2$, $MoSe_2$ and $WSe_2$ flakes with different number of layers. Although this figure could be used as a guide to determine the number of layers of TMDCs exfoliated onto $SiO_2$/Si substrates, the difference in optical contrast spectra between layers with different thicknesses is more subtle than that measured onto the PDMS substrate by differential reflectance. Also the spectra show a skewed 'S' shape because of the interference color effect, due to the thin $SiO_2$ dielectric layer on top of the reflective silicon surface, which hampers the identification of the excitonic features that are superimposed (still visible on the MoS$_2$ flakes, Figure S6a). Therefore, these results illustrate that it is preferable to characterize the TMDCs on the PDMS substrate (by means of the combination of the quantitative analysis of the transmission mode optical images and the differential reflectance/transmittance) prior to their transfer to SiO$_2$/Si substrates.

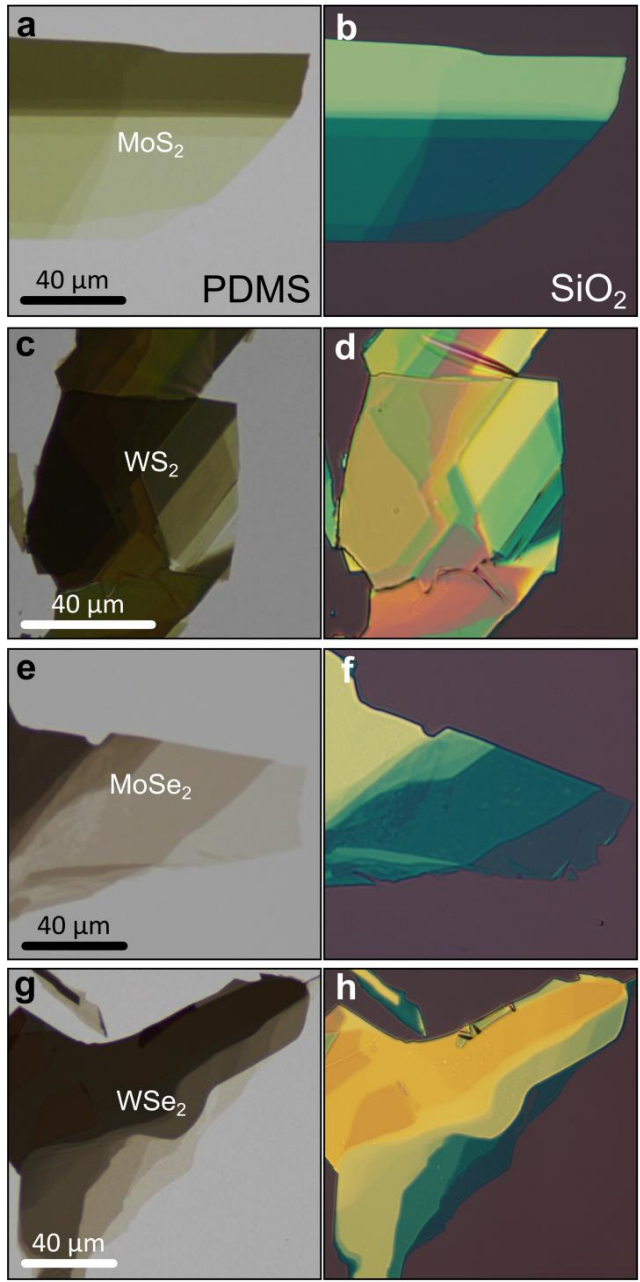

**Figure S5.** Transmission mode optical images (left panels) of MoS$_2$ (a), WS$_2$ (c), MoSe$_2$ (e) and WSe$_2$ (g) on PDMS substrates. Reflection mode optical images of the same flakes after transfer onto SiO$_2$/Si substrates (285 nm thick SiO$_2$): MoS$_2$ (b), WS$_2$ (d), MoSe$_2$ (f) and WSe$_2$ (h). Note: the images on SiO$_2$/Si substrates have been flipped horizontally to facilitate the comparison with the transmission mode images.

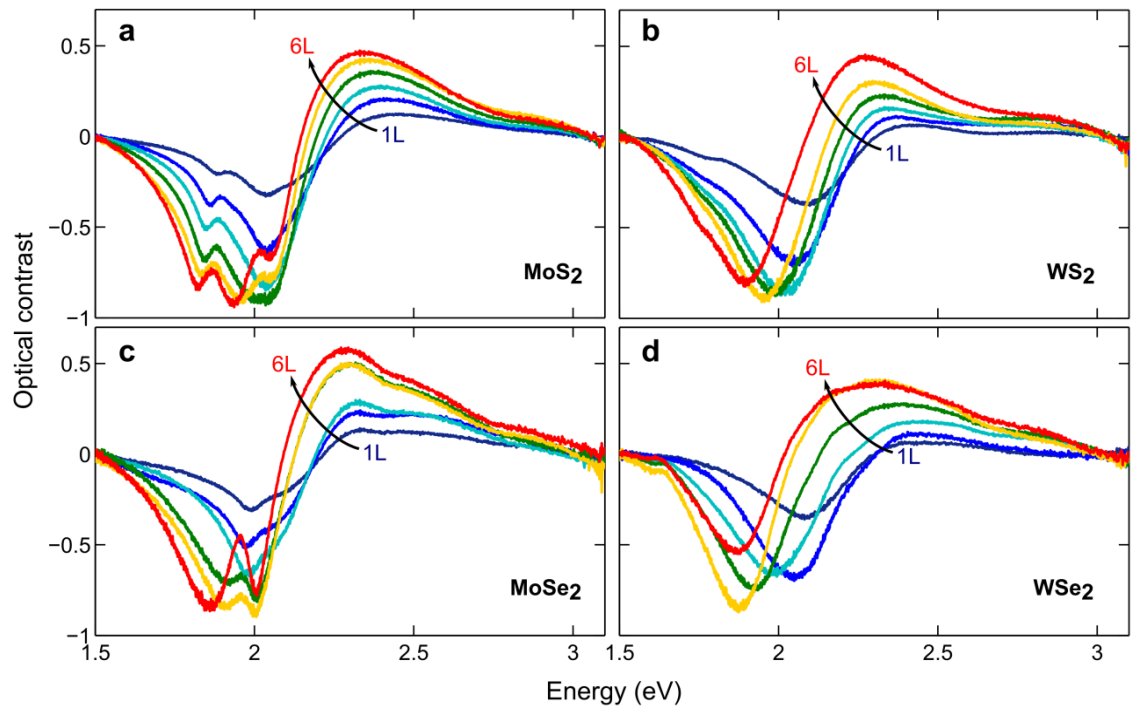

**Figure S6.** Thickness dependence of the optical contrast measured for $MoS_2$ (a), $WS_2$ (b), $MoSe_2$ (c) and $WSe_2$ (d) deposited onto $SiO_2/Si$ substrates (285 nm thick $SiO_2$).

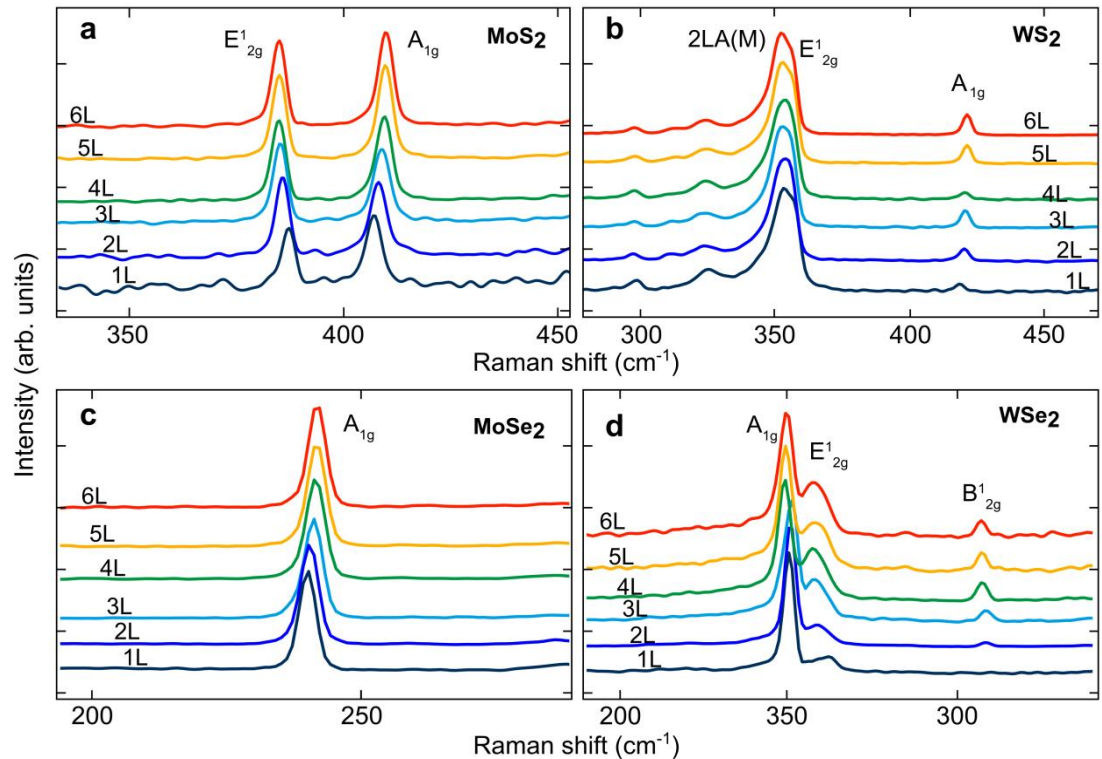

**Figure S7.** Raman spectra mesured on $MoS_2$ (a), $WS_2$ (b), $MoSe_2$ (c) and $WSe_2$ (d) deposited onto PDMS substrates.

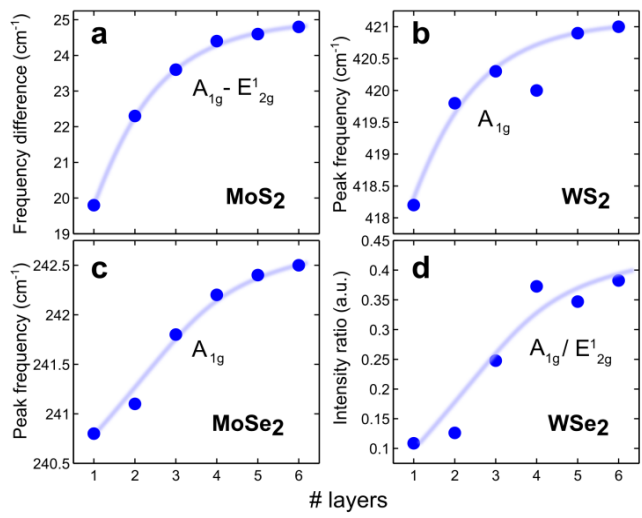

**Figure S8.** Quantitative analysis of the Raman spectra measured on MoS$_2$ (a), WS$_2$ (b), MoSe$_2$ (c) and WSe$_2$ (d) deposited onto PDMS substrates. The solid lines are guides to the eye.

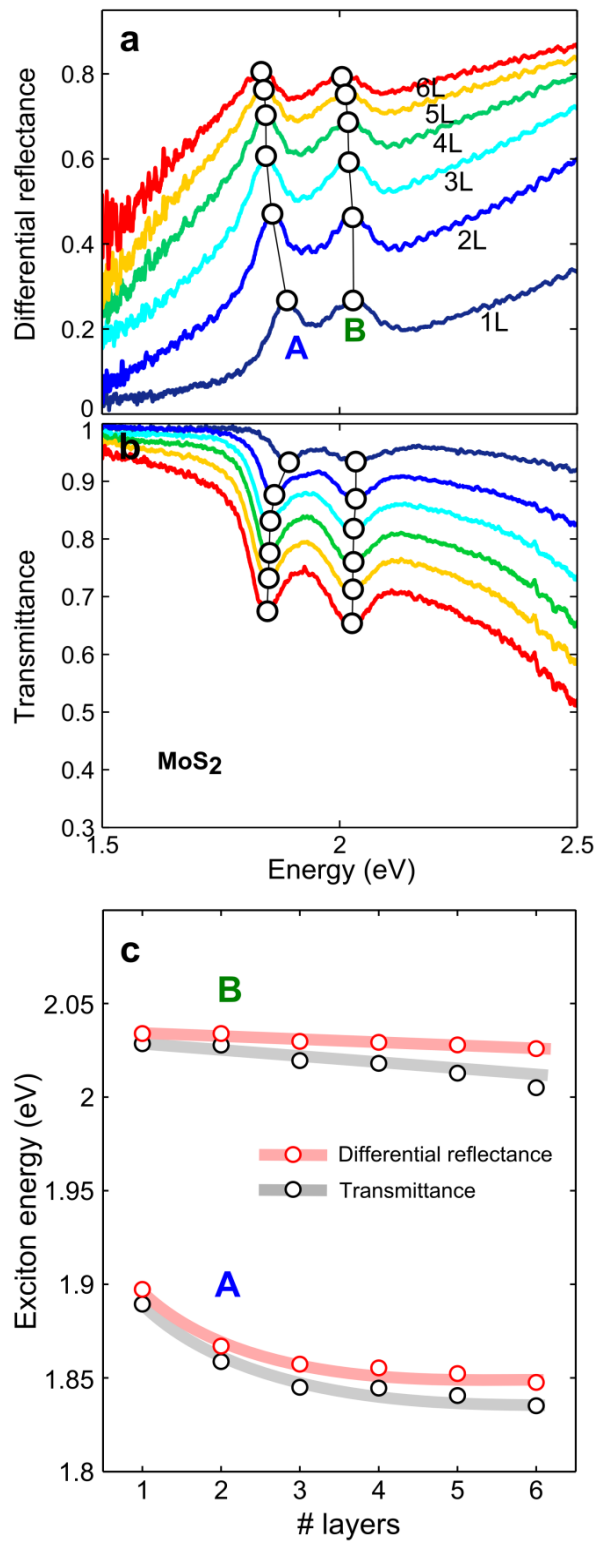

**Figure S9.** Comparison between differential reflectance (a) and transmittance (b) measurements carried out on the same MoS$_2$ flakes on PDMS. (c) Comparison between the exciton energies determined from differential reflectance and transmittance measurements. The slight variation between the two methods could be attributed to a slight increase of temperature of the substrate during the transmittance measurements (leading to a slight biaxial straining of the flakes).

Now we turn our attention to the differential reflectance spectra outside the energy window where the exciton resonances occur. We have found that the differential reflectance magnitude increases monotonically with the number of layers. The results are summarized in Figure 7, demonstrating that a quantitative analysis of epi-illumination images can be used to determine the thickness of the MoS$_2$, MoSe$_2$, WS$_2$ and WSe$_2$ samples. The quantitative analysis could be carried out by selecting the illumination wavelength with narrow-bandpass filters, typically used in most laboratories to enhance the optical contrast of 2D materials.

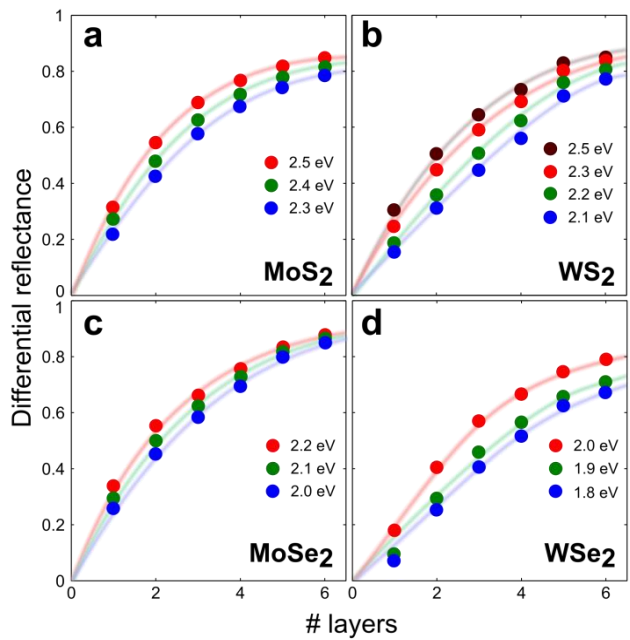

**Figure S10.** Differential reflectance intensity measured from the differential reflectance spectra shown in Figure S6 at energies outside the excitonic resonance windows: (a) MoS$_2$, (b) WS$_2$, (c) MoSe$_2$ and (d) WSe$_2$. The solid lines are guides to the eye.

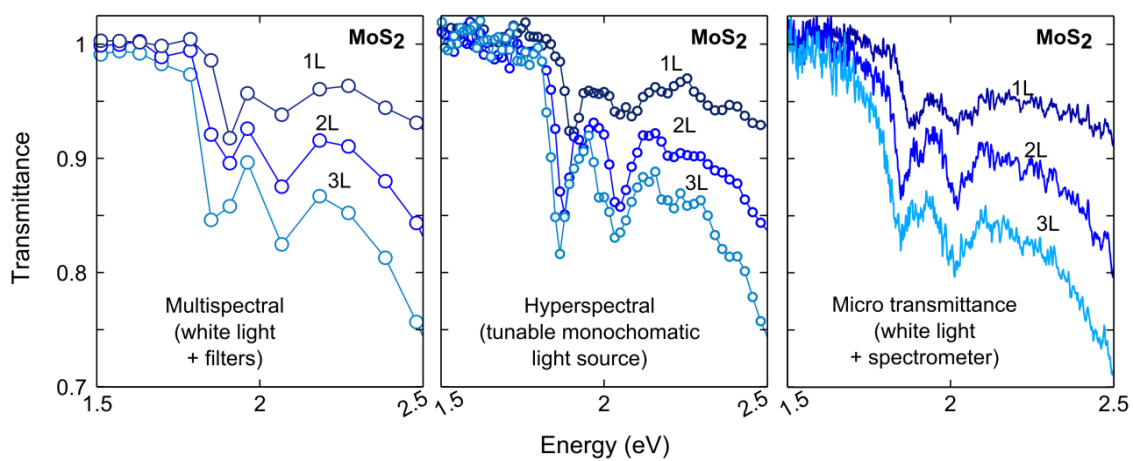

**Figure S11.** Comparison between different methods to measure the optical properties of 2D materials (using 1L, 2L and 3L MoS$_2$ as testbed). In the multispectral measurements narrow bandwidth filters are used to select the illumination wavelength. In the hyperspectral method the illumination is carried out through a white-light source connected to a monochromator. In the micro-transmittance measurement, we employ white light, which is collected through an optical fiber (acting as a confocal pinhole) and sent to a CCD spectrometer.

Figure S12 shows an artistic representation of the crystal structure of 2H- (Figure S11 up) and 3R- (Figure S12 down) MX$_2$ crystals. The 2H and 3R phase differ in the bulk crystals, since it arises from a different stacking of 2D layers, interacting by van der Waals forces. For instance, in MoS$_2$, the 2H phase presents unit cell parameters *a* = *b* = 3.1625 Å and c = 12.300 Å (space group *P6$_3$/mmc*), while the 3R phase presents unit cell parameters *a* = *b* = 3.1607 Å and c = 18.344 Å (space group *R3m*).[1]

The different stacking of the 2H and the 3R phase leads to slightly distinct band structures and, therefore, different excitonic phenomena. In the case of MoS$_2$, for example, the energy splitting of the top of the valence band at the $\bar{K}$ point is smaller for the 3R-MoS$_2$ (0.14 eV) than for the 2H-MoS$_2$ (0.17 eV), which is translated in different exciton splitting [2]. This different splitting can be observed in the differential reflectance spectra (Figure S14 and Figure S15). In Figure S13 we show optical microscopy images of mechanically exfoliated flakes of 2H- and 3R- MoS$_2$ mono- and few-layer crystals on a PDMS substrate, where no difference can be depicted between the two phases. Also, the transmittance extracted from transmission mode images seems very similar for both 2H and 3R polytypes (Figure S14). Therefore, the quantitative analysis of the A and B exciton energy difference seems the more reliable way to distinguish between the 2H and the 3R polytypes.

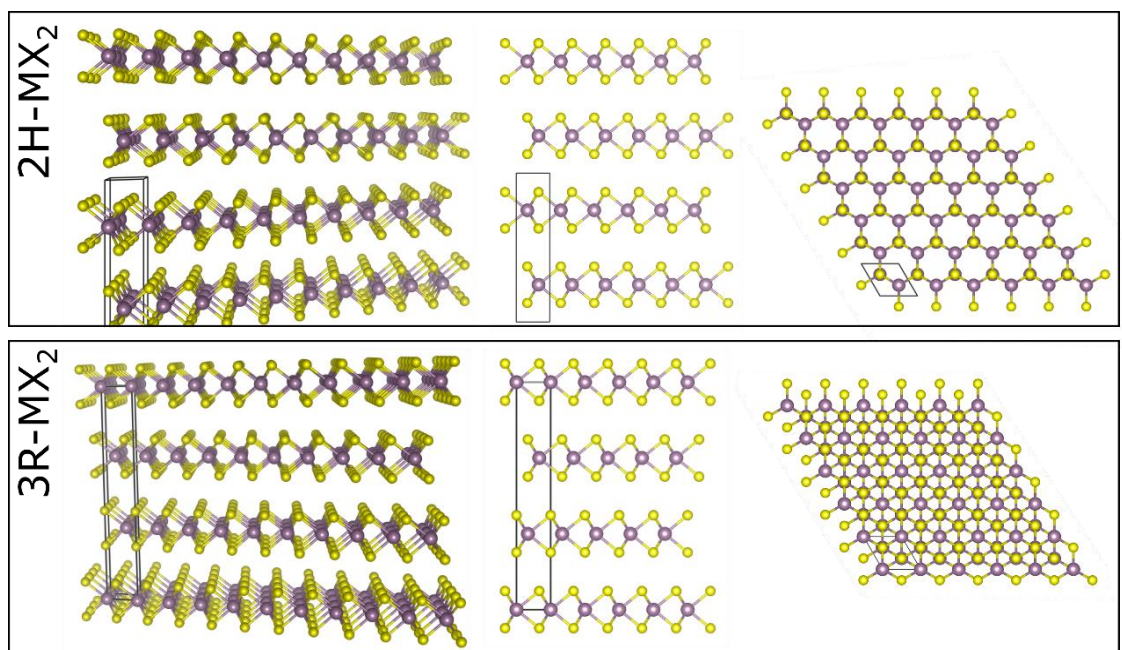

**Figure S12.** Comparison between the crystal structure of the 2H- and 3R- polytypes.

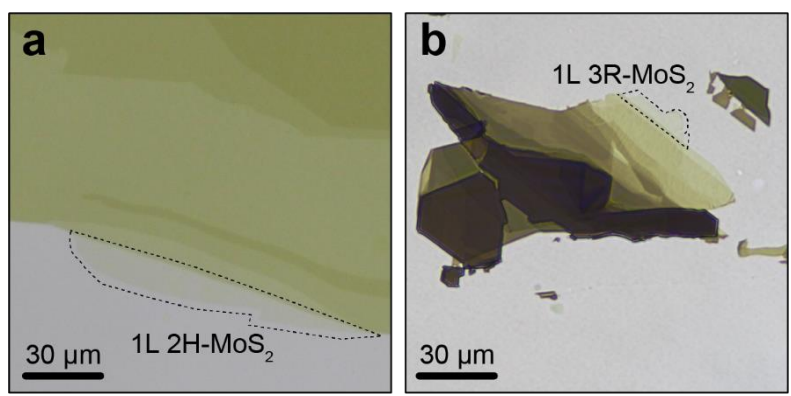

**Figure S13.** Transmission mode optical images of 2H-MoS$_2$ (a) and 3R-MoS$_2$ on PDMS. The dashed regions highlight the single-layer regions. Note that MoS$_2$ single-layers of 2H and 3R have the same structure and they only differ for multilayered stacks.

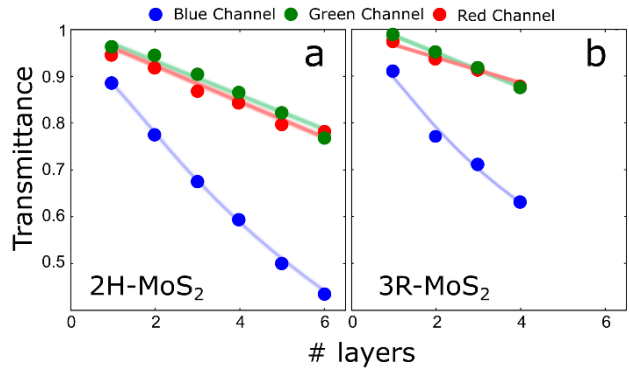

**Figure S14.** Transmittance (extracted from the red, green and blue channels of the trasnmission mode optical images) as a function of the number of layers for (a) 2H-MoS$_2$ (data reproduced from Figure 1d to facilitate the comparison with the 3R polytype) and (b) 3R-MoS$_2$. The solid lines are guides to the eye.

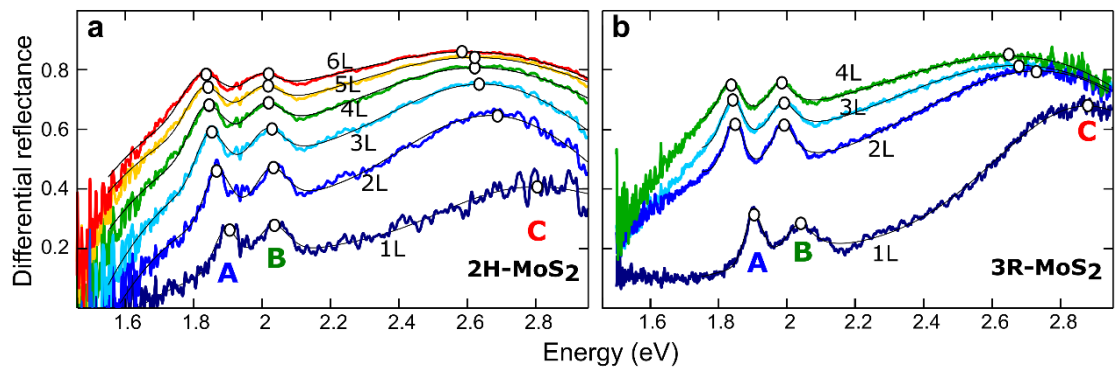

**Figure S15.** Differential reflectance spectra measured as a function of the number of layers for (a) 2H-MoS$_2$ (data reproduced from Figure 1d to facilitate the comparison with the 3R polytype), (b) 3R-MoS$_2$. The spectra have been fitted to a sum of Lorentzian/Gaussian peaks (solid thin black lines) to determine the position of the different excitonic features (highlighted with white circles).

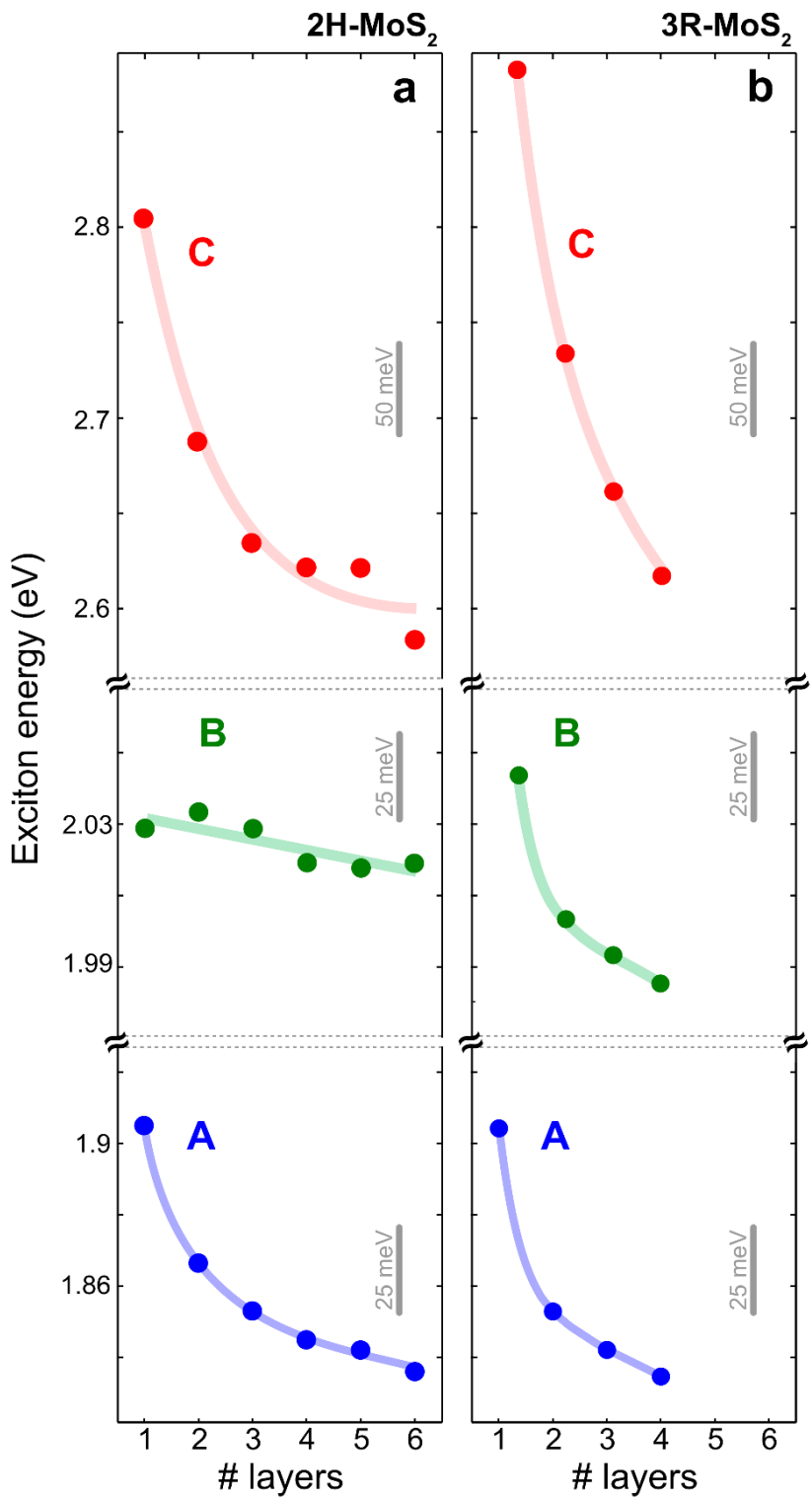

**Figure S16.** Thickness dependence of the exciton energies, extracted from the differential reflectance spectra of (a) 2H-MoS$_2$ (data reproduced from Figure 1d to facilitate the comparison with the 3R polytype) and (b) 3R-MoS$_2$. The solid lines are guides to the eye.

**Details on the *ab initio* calculations:**

All calculations are performed using a code written on our own [3].

To end up with the absorption spectra of the four different TMDCs, we start with a DFT calculation in the LDA approximation using three shells of localized Gaussian orbitals as basis set. Each of the shells is composed of ten orbital functions covering the symmetries s, p, d and s*.

All orbitals inside one shell share the same material dependent decay constant, which are in a range of 0.13 $a_B^{-2}$ to 2.5 $a_B^{-2}$. The reciprocal space is sampled with a 12 × 12 × 1 $k$-point grid for the mono- and bilayers and a 10 × 10 × 3 $k$-point grid for the bulk crystals. We use the structural parameters as reported in Ref. [4] (for $MoS_2$ and $MoSe_2$) and Ref. [5] (for $WS_2$ and $WSe_2$) with experimental lattice constants of 3.160 Å, 3.299 Å, 3.155 Å and 3.286 Å for $MoS_2$, $MoSe_2$, $WS_2$ and $WSe_2$, respectively. The S or Se atoms of the mono- and bilayer system from neighboring unit cells are vertically separated by at least 28 Å vacuum to suppress interactions due to the periodic continuation perpendicular to the layers (in the DFT). Spin-orbit interaction is included in terms of corresponding pseudopotentials and all spin-split bands enter in the consecutive quasiparticle calculation.

The quasiparticle corrections are calculated within the LDA+$GdW$ [6] approximation, which allows for well converged results at comparably low numerical costs. Figure S17 shows the convergence behavior of the direct gaps at the high symmetry point $K$ with respect to the auxiliary plane wave basis to represent ε and $W$. For this convergence study, the $k$-point grid is chosen as 12 × 12 × 1 for the mono- and bilayers and 12 × 12 × 3 for the bulk crystals. The data in Figure S1 show that a plane wave basis of 2.5 Ry (205 plane waves) is already sufficient. At both levels, DFT and $GW$, the spin-orbit interaction is fully taken into account.

In order to get the absorption spectra we solve the Bethe-Salpeter equation (BSE) using identical $k$-point grids for the quasiparticle corrections and the electron-hole interactions therefore avoiding the need of an interpolation scheme. The A exciton is the lowest optically bright excitation. The B exciton corresponds to the next optically bright excitation that is not an excited state of the A exciton. Figure S18 and Figure S19 summarize the convergence of the A and B excitons. Apparently, a $k$-point grid of 24× 24 × 1 (mono-/bilayer) or 18 × 18 × 3 (bulk) yields well-converged results. These $k$-point grids are employed for the data shown in Figure S20 and Figure S21. Since the C (D) exciton is composed of several excitations, we calculate the excitation energy as a weighted sum over all excitations inside an energy window that is chosen such that the leading and tailing edges of the peaks are dropped equally to the level of the absorption background (see Figure S20 for more details). To account for uncertainties in the definition of the C (D) exciton, we introduce an error in the respective energetic positions. For the monolayers we include four valence and six conduction bands, while for the bilayer and bulk crystals these numbers are doubled (since the number of atoms in the unit cell are doubled).

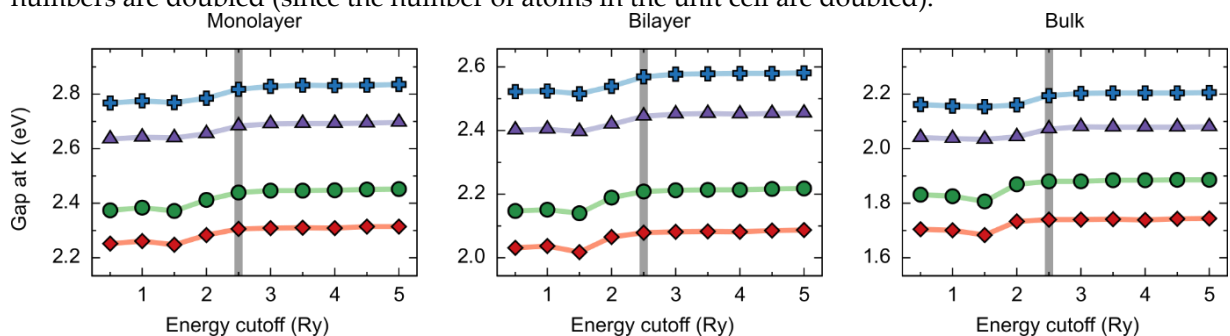

**Figure S17.** Convergence of the quasiparticle gap at the $K$ point with respect to the energy cutoff used in the LDA+$GdW$ approach for the representation of ε and $W$. For the mono- and bilayers a $k$-point grid of 12 × 12 × 1 and for the bulk crystals a $k$-point grid of 12 × 12 × 3 is used. All four materials $MoS_2$ (✚), $MoSe_2$ (●), $WSe_2$ (◆) and $WS_2$ (▲) show similar convergence behaviour for all three numbers of layers. The grey line shows the chosen energy cutoff of 2.5 Ry employed for preparing the subsequent BSE calculations.

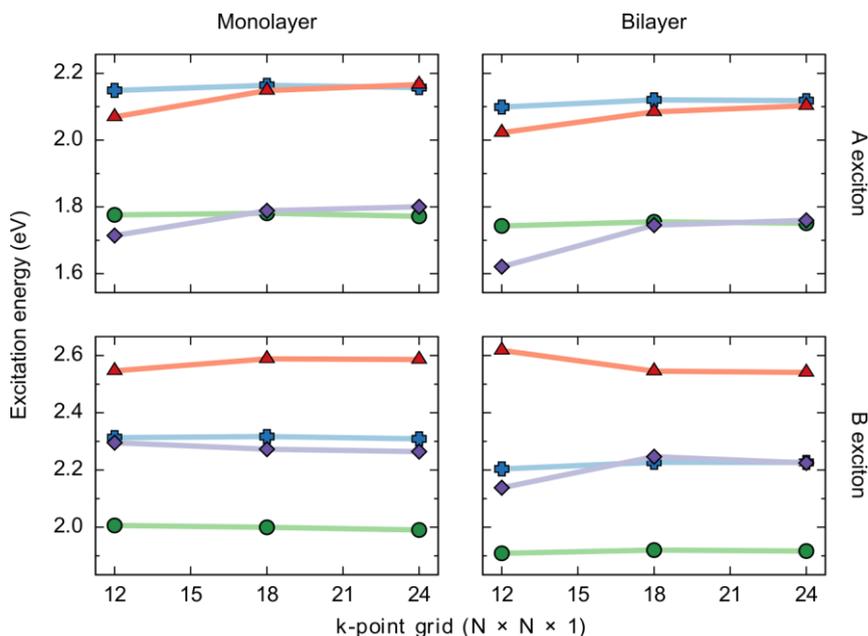

**Figure S18.** Convergence of the A and B exciton for all four TMDCs $MoS_2$ (✚), $MoSe_2$ (●), $WS_2$ (▲) and $WSe_2$ (◆) with respect to the $k$-point grid used in the BSE. For the monolayers four valence and six conduction bands are included and for the bilayers eight valence and twelve conduction bands were taken into account. The solid lines are guides to the eye.

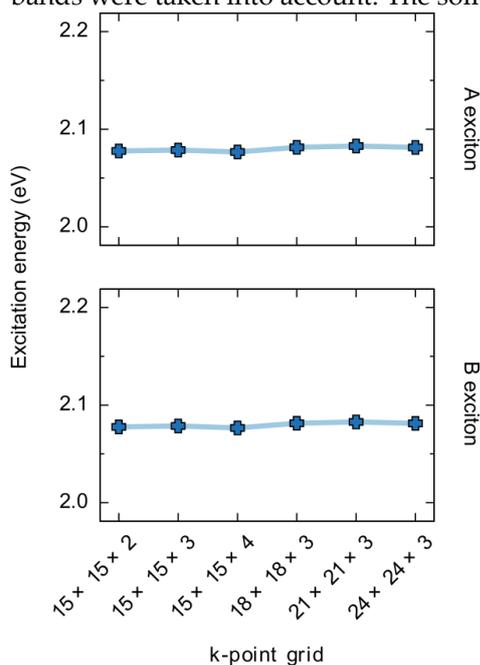

**Figure S19.** Convergence of the A and B exciton for the bulk crystal of $MoS_2$ with respect to the $k$-point grid applied in the BSE. Note that the number of bands were reduced to four valence and six conduction bands for these calculations to facilitate the calculation with $24 \times 24 \times 3$ $k$-points. The solid lines are guides to the eye.

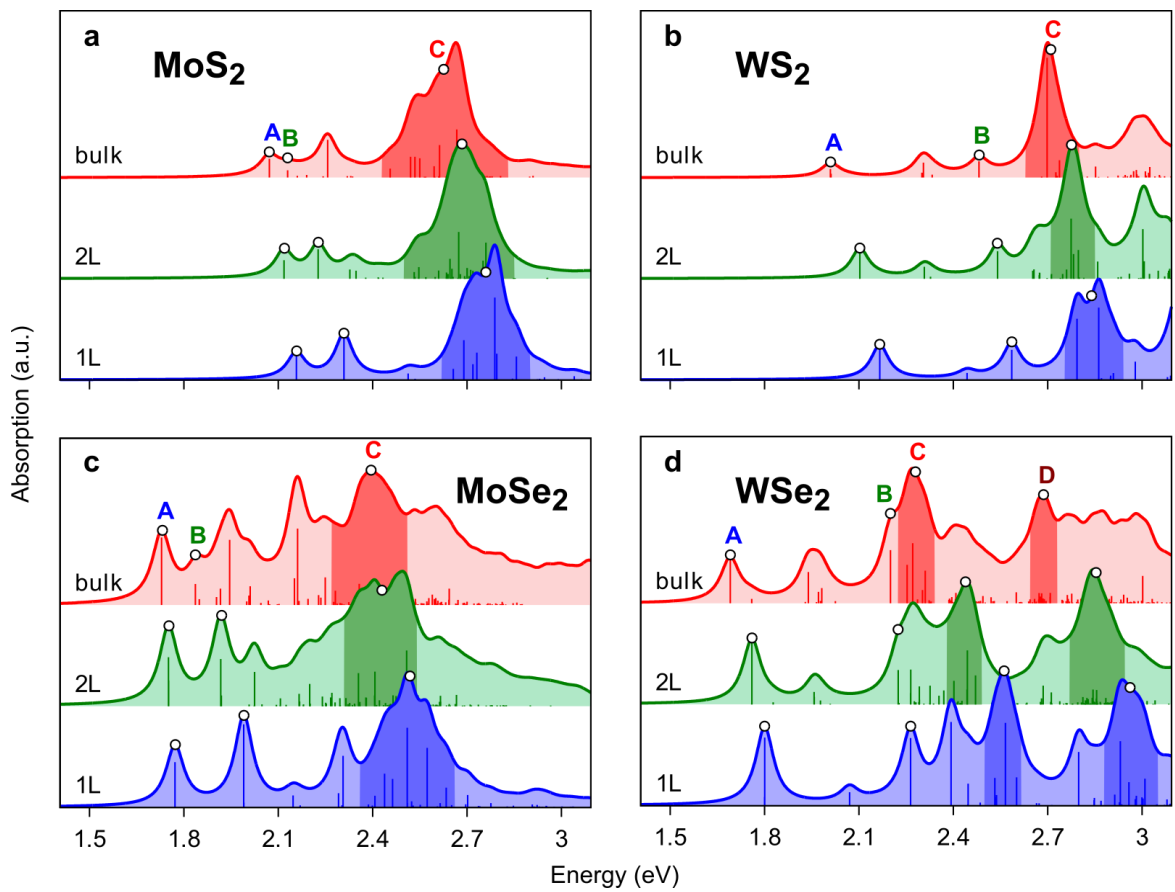

**Figure S20.** Absorption spectra of MoS$_2$ (**a**), WS$_2$ (**b**), MoSe$_2$ (**c**) and WSe$_2$ (**d**) for all studied number of layers. An artificial broadening of 35 meV is introduced and the spectra are vertically shifted to distinguish between the different systems of one material. The energy window chosen for the weighted sum of the respective C and D excitons are marked by the darker regions of the spectra.

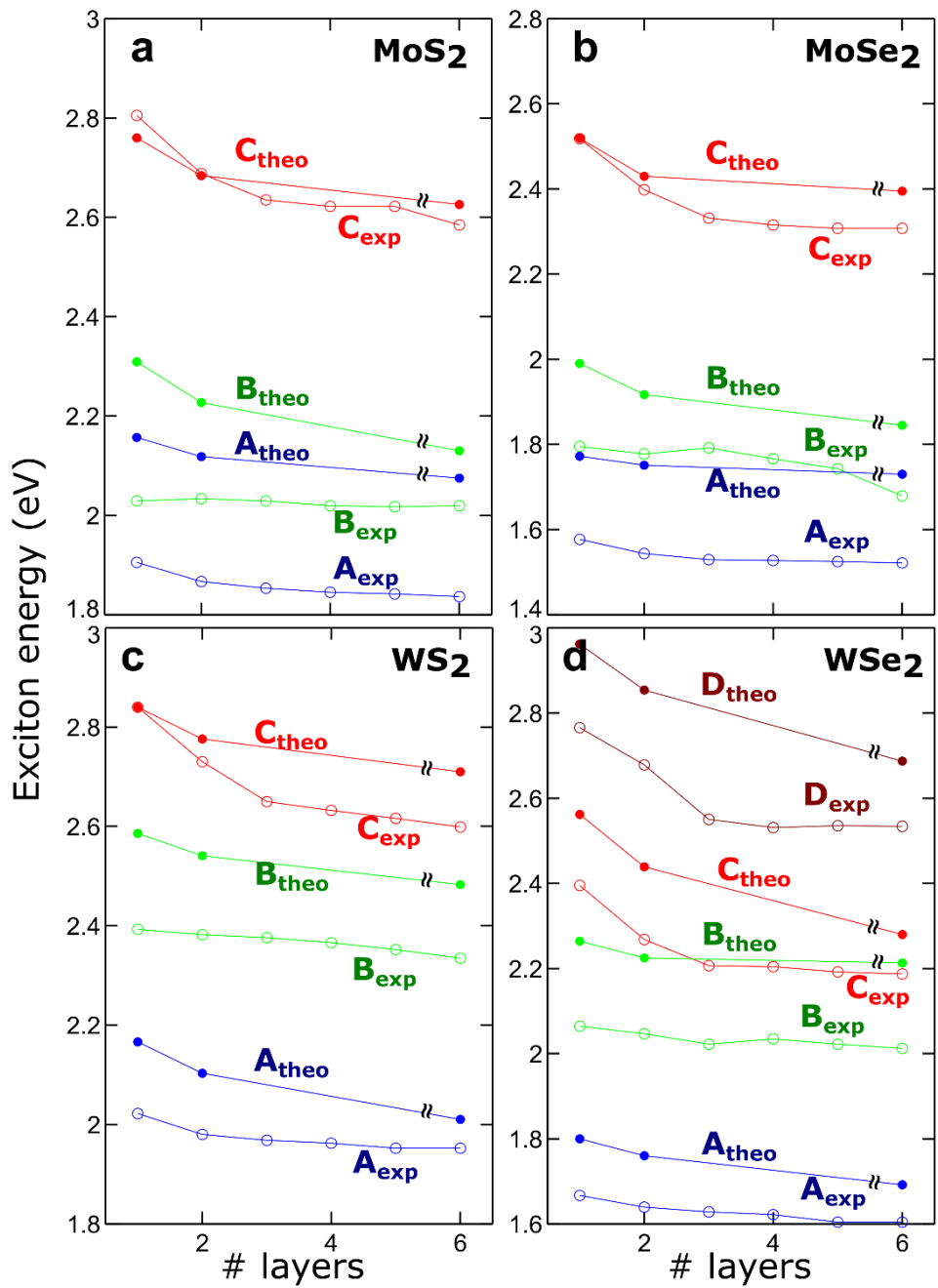

**Figure S21.** Direct comparison between the exciton energies obtained experimentally and those calculated for MoS$_2$ (**a**), WS$_2$ (**b**), MoSe$_2$ (**c**) and WSe$_2$ (**d**).

**Reproducibility of the differentia reflectance spectra**

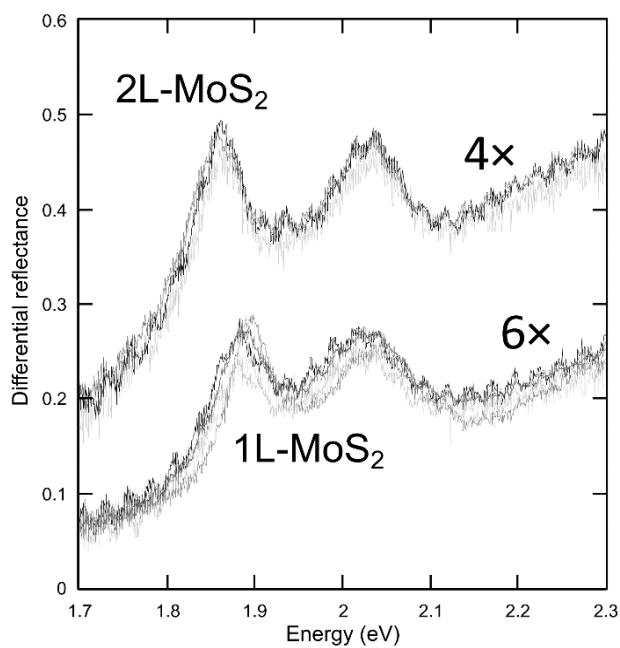

**Figure S22.** Comparison between differential reflectance spectra acquired on different single-layer and bilayer $MoS_2$ flakes that illustrates how the flake-to-flake variation is the main source of uncertainty in the analysis of the thickness dependent spectra of TMDCs. We found a variation of up to 15 meV in the position of the excitons measured on different $MoS_2$ flakes.

**Scheme of the experimental setup**

For a comprehensive description of the experimental setup, its calibration and implementation we address the reader to Reference [27] of the main text.[7]

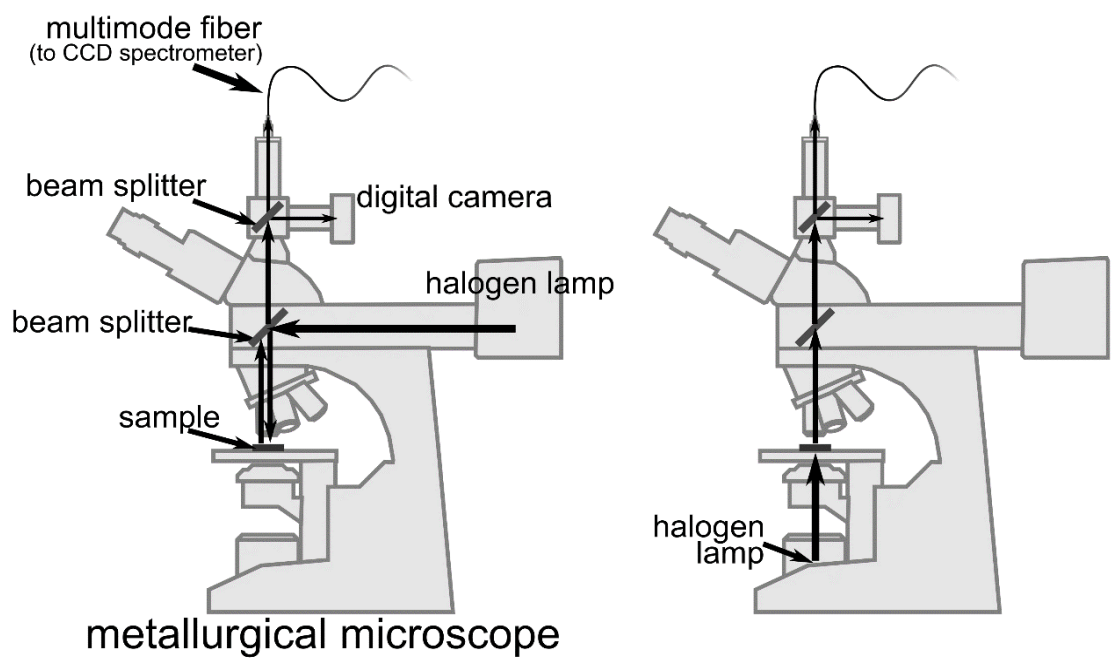

**Figure S23.** Schematic of the experimental setup employed for the micro-reflectance (left) and transmittance (right) measurements on TMDCs. The trinocular of a commercial metallurgical microscope is modified to include an optical path where the image of the sample is projected over the core of a multimode optical fiber, collecting the reflected/transmitted light from a small part of the sample and acting as a confocal pinhole. The other end of the multimode fiber is connected to a compact CCD spectrometer (Thorlabs CCS200/M) to quantitatively analyze the reflected/transmitted light.

**SUPP. INFO. REFERENCES**

(1)    Schönfeld, B., Huang, J. J., & Moss, S. C. Anisotropic mean-square displacements (MSD) in single-crystals of 2H-and 3R-MoS2. *Acta Crystallographica Section B: Structural Science* **1983**, *39*(4), 404-407.

(2)    Suzuki, R., *et al*. Valley-dependent spin polarization in bulk MoS2 with broken inversion symmetry. *Nature Nanotechnology*, **2014** *9*(8), 611-617.

(3)    Rohlfing, M.; Krüger, P.; Pollmann, J. Quasiparticle band-structure calculations for C, Si, Ge, GaAs, and SiC using Gaussian-orbital basis sets. *Phys. Rev. B* **1993**, *48*, 17791-17805.

(4)    Böker, T.; Severin, R.; Müller, A.; Janowitz, C.; Manzke, R.; Voß, D.; Krüger, P.; Mazur, A.; Pollmann, J. Band Structure of MoS$_2$, MoSe$_2$, and $\alpha-$MoTe$_2$: Angle-Resolved Photoelectron Spectroscopy and *ab initio* Calculations. *Phys. Rev. B* **2001**, *64*, 235305.

(5)    Yun, W. S.; Han, S. W.; Hong, S. C.; Kim, I. G.; Lee, J. D. Thickness and Strain Effects on Electronic Structures of Transition Metal Dichalcogenides: 2H-*MX*$_2$ Semiconductors (*M* = Mo, W; *X* = S, Se, Te). *Phys. Rev. B* **2012**, *85*, 33305.

(6)    Rohlfing, M. Electronic Excitations from a Perturbative LDA+*GdW* Approach. *Phys. Rev. B* **2010**, *82*, 205127.

(7) Frisenda, R.; Niu, Y.; Gant, P.; Molina-Mendoza, A. J.; Schmidt, R.; Bratschitsch, R.; Liu, J.; Fu, L.; Dumcenco, D.; Kis, A.; Perez De Lara, D.; Castellanos-Gomez, A. Micro-reflectance and transmittance spectroscopy: a versatile and powerful tool to characterize 2D materials. J. Phys. D. Appl. Phys. 2017, 50, 74002.